\documentclass[1p,sort&compress]{elsarticle}

\usepackage{float}
\usepackage{hyperref}
\usepackage{makeidx}         

\usepackage{amscd,amssymb,amsmath,latexsym,bm}
\usepackage[mathcal,mathscr]{euscript}
\usepackage{lipsum}
\usepackage{amsfonts}
\usepackage{graphicx}
\usepackage{epstopdf}
\usepackage{algorithmic}
\usepackage{hyperref}
\usepackage{xcolor}
\usepackage{enumerate}
\usepackage{ulem}

\numberwithin{equation}{section}
\numberwithin{figure}{section}

\newcommand{\CM}{{\mathbb C}}

\newcommand{\RM}{{\mathbb R}}

\newcommand{\ZM}{{\mathbb Z}}
\newcommand{\PM}{{\mathbb P}}

\newcommand{\GM}{{\mathbb G}}

\newcommand{\EM}{{\mathbb E}}

\newcommand{\Pp}{{\mathcal P}}

\newcommand{\Bb}{{\mathcal B}}
\newcommand{\Dd}{{\mathcal D}}

\newcommand{\Nn}{{\mathcal N}}

\newcommand{\Ll}{{\mathcal L}}

\newcommand{\Hh}{{\mathcal H}}

\begin{document}

\title{Disordered Crystals from First Principles II: Transport Coefficients}

\author{Thomas D. K\"uhne$^{1}$, Julian Heske$^{1}$ and Emil Prodan$^{2}$}

\address{\ $^{1}$ Chair of Theoretical Chemistry, University of Paderborn, Paderborn, Germany
\\$^2$ Department of Physics, Yeshiva University, New York, New York, USA
}

\begin{abstract}
{\scriptsize This is the second part of a project on the foundations of first-principle calculations of the electron transport in crystals at finite temperatures, aiming at a predictive first-principles platform that combines {\it ab-initio} molecular dynamics (AIMD) and a finite-temperature Kubo-formula with dissipation for thermally disordered crystalline phases. The latter are encoded in an ergodic dynamical system $(\Omega,\GM,{\rm d}\PM)$, where $\Omega$ is the configuration space of the atomic degrees of freedom, $\GM$ is the space group acting on $\Omega$ and ${\rm d}\PM$ is the ergodic Gibbs measure relative to the $\GM$-action. We first demonstrate how to pass from the continuum Kohn-Sham theory to a discrete atomic-orbitals based formalism without breaking the covariance of the physical observables w.r.t. $(\Omega,\GM,{\rm d}\PM)$. Then we show how to implement the Kubo-formula, investigate its self-averaging property and derive an optimal finite-volume approximation for it. We also describe a numerical innovation that made possible AIMD simulations with longer orbits and elaborate on the details of our simulations. Lastly, we present numerical results on the transport coefficients of crystal silicon at different temperatures.}
\end{abstract}


\maketitle

\setcounter{tocdepth}{3}

{\scriptsize
\tableofcontents
}
\section{Introduction}

As the scale of the fabrication processes of electronic components is continuously reduced, the quantum mechanical aspects of the charge transport become more important and {\it ab-initio} quantum simulations will be required for an accurate and predictive characterization. Since most electronic components operate at room and higher temperatures, these {\it ab-initio} simulations have to take into account the thermal motion of the atoms. Since the dynamics of the electrons is orders of magnitude faster than that of the ionic cores, the quantum dynamics of the electrons takes place in a highly disordered environment. This can result in qualitatively different dynamical behaviors, notably the absence of quantum diffusion or Anderson localization \cite{AbrahamsPRL1979}, that can not be captured by empirical models or idealistic zero temperature simulations. This and other effects will be investigated from first-principles in this work. To the best of our knowledge, the work reported here is the first attempt to simulating quantum charge transport at finite-temperature from first-principles. 

\vspace{0.2cm}

Most of us think of crystals as condensed phases of matter, where the atoms are periodically arranged in space. However, the crystalline phase persists all the way to the melting point so clearly that oversimplifying picture is highly misleading. In fact, defining the crystalline phase is a deep and highly non-trivial problem in condensed matter physics. On the formalism side, the works of Bellissard on the homogeneous phases of matter represent a milestone \cite{Bellissard2003,Bellissard2015}. They taught us that, in crystals, the space group symmetry $\GM$ manifested at zero temperature is replaced at finite temperatures by an ergodic $\GM$-action w.r.t. the Gibbs measure on the space $\Omega$ of thermally disordered atomic configurations. Furthermore, the invariance w.r.t. $\GM$ of the electronic Hamiltonians manifested at zero temperature is replaced at finite temperatures by the covariance w.r.t. the $\GM$-action. Ergodicity and convariance w.r.t. the $\GM$-action explain why the measurements of the macroscopic physical observables, including the transport coefficients, do not fluctuate from one configuration to another and why the symmetry w.r.t. the full space group is restored at the macroscopic level. For example, the latter is manifested in the $\GM$-symmetric X-ray diffraction patterns observed all the way to the melting point \cite{Fitting1999}. Another manifestation is the stability of the topological phases of matter stabilized by point symmetries in conditions where thermal disorder breaks these symmetries \cite{SongPRB2015}.

\vspace{0.2cm}

In our previous work \cite{KP2018}, we took the task of quantifying the ergodic dynamical system $(\Omega, \GM,\tau,{\rm d}\PM)$ that defines a crystalline phase. Using crystal silicon (Si) as a working example, we devised an algorithm that extracts this data from the output of conventional {\it ab-initio} molecular dynamics (AIMD) simulations \cite{CPMD2, TDK2014}. In particular, we were able to quantify and parametrize the Gibbs measure for crystalline Si at various temperatures. In this work, our focus is mostly on the electronic degrees of freedom, which are simulated with hybrid Gaussian-plane wave based density functional theory (DFT) electronic structure codes \cite{Quickstep}. In the first part of our work, we demonstrate how to generate effective lattice models that encode the entire output of the electronic structure codes and where the covariance w.r.t. $(\Omega, \GM,\tau,{\rm d}\PM)$ is explicitly manifested. Particular attention will be dedicated to the tight-binding expressions of the Kohn-Sham (KS) Hamiltonian, position and charge current operators.

\vspace{0.2cm}

For the charge transport, we adopt the non-commutative Kubo-formula derived by Schulz-Baldes and Bellissard \cite{BellissardJMP1994,SBaldesJSP1998,SBaldesRMP1998}. One extremely important aspect of their formalism is that it includes dissipation. More precisely, given a dissipation mechanism encoded in a scattering operator, the formalism produces a dissipation super-operator that is organically incorporated in the Kubo-formula (see section~\ref{Sec:KF}). Various dissipation mechanisms and their corresponding super-operators have been analyzed in \cite{AndroulakisJSP2012} and they certainly can be evaluated from first principles. One should be aware that dissipation has an important role in shaping the I-V characteristics of both metals and semiconductors and this is why a Kubo-formula that incorporates dissipation is so valuable. 

\vspace{0.2cm}

The Kubo-formula derived by Schulz-Baldes and Bellissard has been numerically implemented in the past for disordered tight-binding model Hamiltonians \cite{ProdanAMRX2013,XuePRB2012,XuePRB2013,SongEPL2014,
ProdanAOP2016,ProdanSpringer2017} and other types of aperiodic Hamiltonians \cite{CancesJMP2017,MassattMMS2018,MassattEPJP2020}. One of the main findings of these works is the rapid convergence of the results with system size. For example, in systems with known quantized transport coefficients, such as 2-dimensional Hall systems, the non-commutative Kubo-formula reproduced the quantization with two digits of precision even on small $10 \times 10$ highly disordered tight-binding lattices. This is a convincing fact that this approach is highly suited for the applications we seek in this work, given that the super-cells that can be handled by first-principles simulations are inherently small.  

\vspace{0.2cm}

The simulations we report here for crystalline Si at different temperatures are preliminary and certainly not converged w.r.t. either the system's size, or the atomic orbital basis, but they are certainly converged w.r.t. the thermal disorder sampling. Also, the dissipation super-operator is treated in the relaxation time approximation where it becomes proportional to the identity map. The simulations produced expected outputs for the available electronic structures and enabled us to test several important qualitative aspects of the charge transport. One aspect is the formation of a dynamical band gap where the quantum diffusion is absent and this dynamical gap was found to be much larger than the spectral gap. The former defines the reference for the activated behavior of the conductivity, while the latter for the charge carriers. Since these are two different reference energy levels, the Anderson localization phenomenon can lead to substantial quantitative effects that were overlooked so far. We also found that the conductivity tensor is extremely sensitive to the dissipation relaxation time. Given this sensitivity, we believe that the prevalent dissipation mechanism in crystalline Si at room temperature can be identified with high precision by future simulations which incorporate first principles dissipation super-operators.

\vspace{0.2cm}

Based on previous tight-binding model simulations, we initially estimated that at least 1000 disordered atomic configurations will be necessary and, as such, we performed large time scale AIMD simulations to acquire that amount of data. However, our calculations revealed that the average over the atomic configurations of the conductivity tensor can be achieved with a relatively small number of configurations, which can be as low as 50. In fact, with reasonable level of dissipation, the thermal fluctuations are almost entirely suppressed for the largest crystal we simulated, which is a direct manifestation of the self-averaging property of the Kubo-formula. This finding assures us that, in the future simulations, we can reduce the time scale of AIMD simulations, hence, enabling us to further increase the crystal size and to better optimize the orbital basis.

\section{Thermal Disorder from First Principles}
\label{Sec:Recap}

In order to fix our notations and provide the context for the present calculations, we briefly recall our main results reported in \cite{KP2018}. Therein, we describe the ergodic dynamical system $(\Omega, \GM,\tau,{\rm d}\PM)$, which completely characterizes the crystalline phase of Si at finite-temperature, where $\Omega$ is the atomic configuration space, $\GM$ is the space group, ${\rm d}\PM$ is the Gibbs measure and $\tau$ is an ergodic action of $\GM$ on $\Omega$.

\subsection{The ideal lattice and its symmetries}

\vspace{0.2cm}

\begin{figure}[ht!]
\center
  \includegraphics[width=0.6\textwidth]{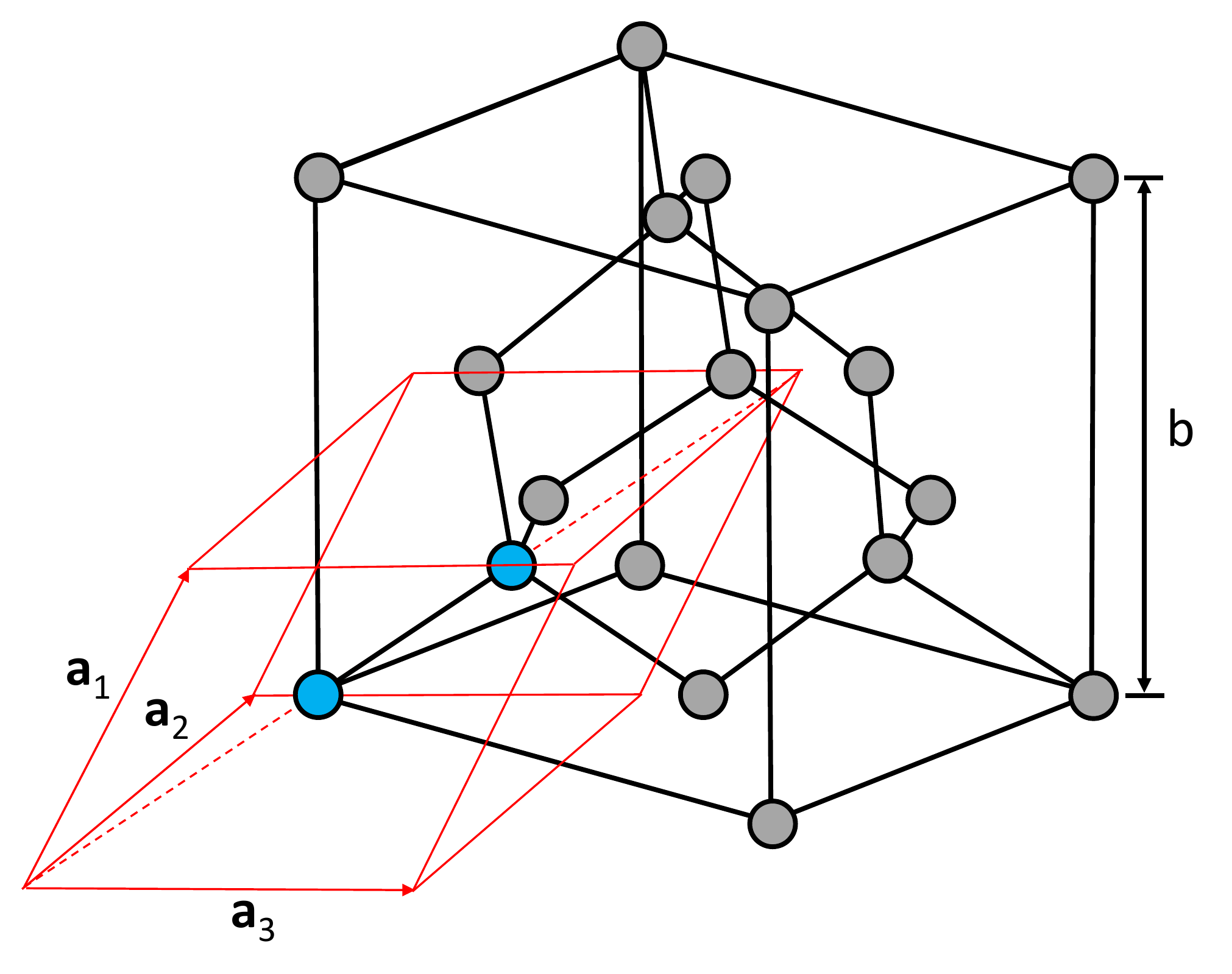}\\
  \caption{\small Si crystalizes in a diamond cubic lattice (Fd-3m), whose conventional unit cell is shown in this diagram. This cubic unit cell is symmetric to the full point group and contains eight Si atoms. The diamond cubic lattice results from the inter-penetration of two face-centered cubic (fcc) lattices. The fcc lattice can be generated by translating a primitive cell that contains just one atom. Hence, silicon's diamond cubic lattice can be generated by translating the same primitive cell, but with one additional Si atom inside it. This primitive cell is shown in red, together with its two atoms  (blue disks) and the generating primitive vectors $\bm a_1=\tfrac{b}{2}(\hat z + \hat y)$, $\bm a_2=\tfrac{b}{2}(\hat x + \hat z)$ and $\bm a_3=\tfrac{b}{2}(\hat y + \hat x)$, with $b=5.431$~\AA. The magnitude of the primitive vectors is $a=b/\sqrt{2}$.}
 \label{SiPrimitiveCell}
\end{figure}

The crystal structure of Si is summarized in Fig.~\ref{SiPrimitiveCell}. Its space group is $\GM=Fd\bar 3$m \cite{Singh93,IntTables}, whose structure is summarized by the following exact sequence of groups
\begin{equation}\label{Eq:GroupSequence}
1 \rightarrow \Bb \rightarrow \GM \rightarrow \Pp \rightarrow 1,
\end{equation}
capturing the extension of the point group $\Pp \subset O(3)$ by the group of discrete translations $\Bb$. The latter can be pictured as the Bravais lattice of the crystal (hence our notation $\Bb$), {\it i.e.} the discrete sub-group of $\RM^3$ defined by the centers of the primitive cells
\begin{equation} 
\Bb = \big \{ n_1 \bm a_1 + n_2 \bm a_2 + n_3 \bm a_3, \ \bm n=(n_1,n_2,n_3) \in \ZM^3 \big \},
\end{equation}
with the generators $\bm a_i$ supplied in Fig.~\ref{SiPrimitiveCell}. The point group $\Pp$ of crystalline Si is O$_h^7$, the full symmetry group of the cube.

\vspace{0.2cm}

Let us recall that a space group is called symmorphic if the exact sequence \eqref{Eq:GroupSequence} is split. Silicon's cubic-diamond lattice is an example of a non-symmorphic space group. Nevertheless, every element $\mathfrak g$ of $\GM$ can be presented in the form $\mathfrak g = (\mathfrak p|\mathfrak a)$, with $\mathfrak p \in \Pp$ and $\mathfrak a \in \RM^3$. Note that for a symmorphic space group, $\mathfrak a$ can be always drawn from $\Bb$, but this is not the case here. Such space group elements act on the points of the Euclidean space as
\begin{equation}
(\mathfrak p|\mathfrak a) \bm x = \mathfrak p \bm x + \mathfrak a, \quad \bm x \in \RM^3.
\end{equation}
They also act on any subset $\Ll$ of the Euclidean space, such as a lattice, via
\begin{equation}
\mathfrak g \Ll = \{\mathfrak g \bm x, \ \bm x \in \Ll\}.
\end{equation} 
The multiplication of the elements takes the form
\begin{equation}\label{Eq:GMultiplication}
 (\mathfrak p|\mathfrak a)(\mathfrak p'|\mathfrak a') =(\mathfrak p \mathfrak p'|\mathfrak p \mathfrak a' + \mathfrak a)
 \end{equation}
 and the inverse of an element is
\begin{equation}
(\mathfrak p|\mathfrak a)^{-1} = (\mathfrak p^{-1}|-\mathfrak p^{-1}\mathfrak a).
\end{equation}

\vspace{0.2cm}

The ideal or zero temperature Si lattice will be denoted by $\Ll_0$. This lattice is left invariant by the space group $\GM$. In fact, the asymmetric unit cell of the diamond cubic structure contains a single atom \cite{Singh93,IntTables}, which means that the entire lattice can be reconstructed from one single point by acting with the full space group: $\Ll_0=\big \{ \mathfrak g \cdot \bm x_0, \ \bm g \in \GM\big \}$. While $\bm x_0$ can be any point of the Euclidean space, we will fix $\bm x_0$ at the origin.

\subsection{Thermally disordered configurations}
\label{Sec:ThDisorder}

\vspace{0.2cm}

When the temperature is finite, the Si atoms undergo a thermal motion and the instantaneous snapshots of the Si lattice can be labeled by a configuration space $\Omega$. In \cite{KP2018}, $\Omega$ was found to be well represented by a product of identical balls
\begin{equation}
\Omega = \prod_{\mathfrak g \in \GM} \, B_{\mathfrak g}, \quad B_{\mathfrak g}=B_{0},
\end{equation}
where a point $\omega = \{\omega_{\mathfrak g}\}_{\mathfrak g \in \GM}$ of $\Omega$ encodes the displacements of the atoms from their equilibrium positions. Thermal motion defines an ergodic dynamical system $\omega(t)$ ($t=$ time) over $\Omega$ and an instantaneous snapshot of crystalline Si supplies a thermally disordered lattice
\begin{equation}
\Ll_\omega = \big \{\mathfrak g \bm x_0 + \omega_{\mathfrak g}, \ \mathfrak g \in \GM \big \} \subset \RM^3.
\end{equation}
For these disordered lattices, the invariance of $\Ll_0$ under the space group is replaced by the covariance relation
\begin{equation}
\mathfrak g \Ll_\omega = \Ll_{\tau_{\mathfrak g}\omega}, \quad \forall \ \omega \in \Omega, \quad \mathfrak g \in \GM.
\end{equation}
The action $\tau$ of the space group on $\Omega$, appearing above, can be computed as follows. If $\mathfrak g = (\mathfrak p|\mathfrak a) \in \GM$, then
\begin{align}
\mathfrak g \Ll_\omega & = \{\mathfrak g(\mathfrak g' \bm x_0 +\omega_{\mathfrak g'}), \ \mathfrak g' \in \GM\} \\ \nonumber 
& = \{\mathfrak p(\mathfrak g' \bm x_0) +\mathfrak p \omega_{\mathfrak g'} + \mathfrak a, \ \mathfrak g' \in \GM\}.
\end{align} 
After regrouping,
\begin{equation}
\mathfrak p(\mathfrak g' \bm x_0) +\mathfrak p \omega_{\mathfrak g'} + \mathfrak a = \mathfrak {gg'} \bm x_0 +  \mathfrak p \omega_{\mathfrak g'},
\end{equation}
and, after the change of variable $\mathfrak g' \rightarrow \mathfrak g^{-1} \mathfrak g'$, we have
\begin{equation}
\mathfrak g \Ll_\omega = \{\mathfrak g' \bm x_0 +\mathfrak p \omega_{\mathfrak g^{-1}\mathfrak g'}, \ \mathfrak g' \in \GM\} = \Ll_{\tau_{\mathfrak g} \omega}.
\end{equation}  
We now can identify the action as
\begin{equation}
\tau_{\mathfrak g} \omega = \tau_{\mathfrak g}\{\omega_{\mathfrak g'}\}_{\mathfrak g' \in \GM}=\{\omega'_{\mathfrak g'}\}_{\mathfrak g' \in \GM}, \quad \omega'_{\mathfrak g'}=\mathfrak p \omega_{\mathfrak g^{-1}\mathfrak g'}.
\end{equation}
One can verify that $\tau_{\mathfrak g_1} \tau_{\mathfrak g_2} = \tau_{\mathfrak g_1 \mathfrak g_2}$, as it should be for a group action.

\vspace{0.2cm}

The Gibbs measure over the configuration space $\Omega$ can be computed from the atomic orbits in an AIMD simulation, more precisely, from the histograms encoding the number of times an orbit intersects the elementary volumes of $\Omega$. In \cite{KP2018}, the Gibbs measure of the Si crystal was found to be extremely well characterized by a multivariate normal distribution of zero mean
\begin{equation}
{\rm d}\PM(\omega) = \rho(\omega) {\rm d}\omega, \quad \rho(\omega) = \tfrac{1}{\sqrt{{\rm Det}(2\pi \hat \Sigma)}}e^{-\frac{1}{2} \omega^T \hat \Sigma^{-1} \omega},
\end{equation}
where $\omega$ is seen here as a $1$-column matrix and the variance matrix $\hat \Sigma$ was quantified in \cite{KP2018} as a function of temperature. The Gibbs measure is invariant and ergodic w.r.t. the $\tau$-action. In fact, the Gibbs measure found in \cite{KP2018} is ergodic relative to the subgroup $\Bb$ of translations,  which is in fact a generic property of homogeneous systems at thermodynamic equilibrium \cite[Chap.~6]{RuelleBook}. The crystalline phase of Si at finite-temperature is entirely defined by the ergodic dynamical system $(\Omega, \GM,\tau,{\rm d}\PM)$.

\vspace{0.2cm}

The observations of the last paragraph will play an important role for the self-averaging properties of the transport coefficients. Let us stress again that, due to the well separated scales in the dynamics of the atomic and electronic degrees of freedom, the quantum state of the electrons evolves in a static atomic potential.\footnote{The electron-phonon and electron-electron scattering processes are rare and sudden dynamical events, which are included via Poisson processes as explained later.} This is a thermally disordered potential and, as we shall see, the physical observables, such as the Hamiltonians or charge currents, become indexed by points of the configuration space $\Omega$. In this new context, the notion of a symmetric observable is replaced by that of a covariant observable. The macroscopic measurements of these observables, however, are independent of the thermally disordered configuration. This remarkable property is a consequence of the covariance and of the ergodic character of the Gibbs measure.

\section{Tight-Binding Form of the Physical Observables}

Our goal for this section is to formulate discrete representations of the Hamiltonians and other physical observables in the context of Gaussian-based implementations of the KS program. Special attention will be given to the transformation of the physical observables under the space group $\GM$ of the crystal.

\subsection{The continuum theory}
\label{Sec:ContTh}

\vspace{0.2cm}

The formally exact KS theory for condensed matter systems, at its most fundamental level, is formulated over the Hilbert space $L^2(\RM^3)$ of square integrable wave functions \cite{KohnPR1965}. For the Si crystal in a thermally disordered configuration $\omega$, the KS-Hamiltonian takes the form
\begin{equation}\label{Eq:KSHam}
H_{\rm KS}^\omega = -\frac{\hbar^2}{2m} \bm \nabla^2_{\bm r} - \sum_{\bm x \in \Ll_\omega} \frac{Ze^2}{|\bm r - \bm x|} + V_{\rm xc}[n_\omega](\bm r),
\end{equation}
where $V_{\rm xc}$ is a local potential encoding the exchange and correlation (XC) effects. The latter has a functional dependence on the electron density $n_\omega(\bm r)$, which is to be determined self-consistently. As the notation suggests, the electron density has a dependence on the atomic configuration $\omega$. In fact, this becomes even more apparent if we reformulate \eqref{Eq:KSHam} as a fixed point problem 
\begin{equation}\label{Eq:EDensity}
n_\omega(\bm r) = \langle \bm r | \Phi_{\rm FD}\big(H_{\rm KS}^\omega;T,\mu\big)|\bm r \rangle,
\end{equation}
where $\Phi_{\rm FD}$ is the Fermi-Dirac distribution at temperature $T$ and chemical potential $\mu$ \cite{ProdanJSP2013}. In the following, we assume that this equation has a unique solution for almost all thermally disordered configuration (see \cite{Cances2012} and \cite{Lahbabi2013}).

\vspace{0.2cm}

The starting point of our study is the covariant property of the KS-Hamiltonian under the space group transformations. To understand the origin of this property, we need to go all the way to the Euclidean group $\EM$ of transformations and recall that the XC potential enjoys the following property
\begin{equation}
V_{\rm xc}[n\circ \mathfrak e](\mathfrak e^{-1} \bm r) = V_{\rm xc}[n](\bm r), \quad \mathfrak e \in \EM,
\end{equation}
for any density function $n$ and point $\bm r \in \RM^3$, which can be inferred from the universality and uniqueness assumptions on $V_{\rm xc}$ \cite{Kohn1965}. Certainly, this can be verified directly for the local density approximation (LDA) to $V_{\rm xc}$ \cite{DMC}. The action of $\EM$ on $\RM^3$  lifts to a unitary action on the Hilbert space $L^2(\RM^2)$ via
\begin{equation}\label{Eq:EAction}
\big ( T_{\mathfrak e}\psi \big )(\bm r) = \psi(\mathfrak e^{-1} \bm r), \quad \mathfrak e \in \EM, \quad \psi \in L^2(\RM^3).
\end{equation}
Now, recall that $\GM$ is just a subgroup of the Euclidean group, hence \eqref{Eq:EAction} describes the action of $\GM$ as well. Then, under such unitary actions, the KS-Hamiltonian behaves as
\begin{align}
 T_{\mathfrak g} H_{\rm KS}^\omega T_{\mathfrak g}^\dagger & = -\frac{\hbar^2}{2m} \bm \nabla^2_{\mathfrak g^{-1}\bm r} - \sum_{\bm x \in \Ll_{\omega}} \frac{Ze^2}{|\mathfrak g^{-1}\bm r - \bm x|} + V_{\rm xc}[n_\omega](\mathfrak g^{-1}\bm r) \\
 & = -\frac{\hbar^2}{2m} \bm \nabla^2_{\bm r} - \sum_{\bm x \in \Ll_{\mathfrak g \omega}} \frac{Ze^2}{|\bm r - \bm x|} + V_{\rm xc}[n_\omega \circ \mathfrak g^{-1}](\bm r).
 \end{align}
 We learn from here that, if $n_\omega$ is the solution of \eqref{Eq:EDensity} for configuration $\omega$, then $n_\omega \circ \mathfrak g^{-1}$ is the solution of \eqref{Eq:EDensity} for configuration $\mathfrak g \omega$. In other words, the self-consistent solutions of the KS equations enjoy the covariant property
 \begin{equation}
 n_{\mathfrak g \omega} = n_\omega \circ \mathfrak g^{-1}.
 \end{equation}
In turn, this assures us that the converged KS-Hamiltonian satisfies the covariant relation
\begin{equation}\label{Eq:KSCov}
T_{\mathfrak g} H_{\rm KS}^\omega T_{\mathfrak g}^\dagger = H_{\rm KS}^{\tau_{\mathfrak g} \omega}, \quad \forall \ \mathfrak g \in \GM.
\end{equation}
It will be extremely important to preserve this characteristics in our tight-binding approximation. As we already mentioned, \eqref{Eq:KSCov} together with the ergodicity of the space group action ensure the self-averaging of the transport coefficients.

\subsection{The effective Hilbert space}

\vspace{0.2cm}

In Gaussian-based approaches, the atom located at position $\bm x \in \Ll_\omega$ carries a finite-dimensional local Hilbert space
\begin{equation}\label{Eq:LocalHilbert}
\Hh_{\bm x} = {\rm Span}\big \{ \phi_n(\bm r-\bm x), \ n=1,\ldots N\big \},
\end{equation}
where $\phi_n :\RM^3 \rightarrow \CM$ are optimized atomic orbitals (see section \ref{Sec:NumSpec} for details). It is important to realize that the same set of functions $\phi_n$ are used for all $\bm x \in \Ll_\omega$. The total Hilbert space for the Gaussian-based computations is the linear subspace
\begin{equation}\label{Eq:Hilbert}
\Hh_\omega =\overline{ {\rm Span} }\big \{ \Hh_{\bm x}, \ \bm x \in \Ll_\omega \big \} \subset L^2(\RM^3).
\end{equation}
As the notation suggests, this subspace depends on the configuration  $\omega \in \Omega$ of the atoms. As we shall see, it is isomorphic to the tight-binding Hilbert space
\begin{equation}
\CM^N \otimes \ell^2(\Ll_\omega) = \overline{\rm Span}\big \{\xi \otimes |\bm x \rangle, \ \xi \in \CM^N, \ \bm x \in \Ll_\omega \big \}
\end{equation} 
of square summable linear combinations of $\xi \otimes |\bm x \rangle$ basis vectors. The scalar product for this space is defined by the orthonormality condition
\begin{equation}
\langle \bm x | \bm x' \rangle = \delta_{\bm x,\bm x'}, \quad \forall \ \bm x,\bm x' \in \Ll_\omega.
\end{equation} 
All our physical observables will be mapped over this tight-binding Hilbert space and all the calculations will be ultimately performed on $\CM^N \otimes \ell^2(\Ll_\omega)$. 

\vspace{0.2cm}

Our goal for this section is to explain in details how to transfer the observables between the Hilbert spaces. We start with the consideration of the overlap coefficients
\begin{equation}\label{Eq:Overlap}
S_{\bm x \bm x'}^{ij}(\omega)= \int_{\RM^3}{\rm d}^3\bm r \ \phi_i^\ast(\bm r -\bm x) \phi_j(\bm r - \bm x'),
\end{equation}
which can be found among the outputs of standard AIMD simulations. Using these coefficients, we form the self-adjoint, positive and invertible operator
\begin{align}\label{Eq:OverlapOp}
& S_\omega: \CM^N \otimes \ell^2(\Ll_\omega) \rightarrow \CM^N \otimes \ell^2(\Ll_\omega), \\ \nonumber
& S_\omega = \sum_{\bm x, \bm x' \in \Ll} \widehat S_{\bm x, \bm x'}(\omega) \otimes |\bm x \rangle \langle \bm x' |,
\end{align}
where $\widehat S_{\bm x, \bm x'}(\omega)$ is the overlap matrix with the entries $S_{\bm x, \bm x'}^{ij}(\omega)$ defined in \eqref{Eq:Overlap}. Then, the isomorphism between $\Hh_\omega$ and $\CM^N \otimes \ell^2(\Ll_\omega)$ is supplied by the unique linear map $U_\omega$ that acts on the generators as
\begin{equation}\label{Eq:Map}
\phi_n(\bm r -\bm x) \mapsto \sqrt{S_\omega} \ \xi_n \otimes |\bm x\rangle, \quad n = 1,\ldots N.
\end{equation}
Above, $\xi_n \in \CM^N$ is a column vector, whose entries are one at position $n$ and zero for all others, whereas $\sqrt{S_\omega}$ is the square root operator defined via the functional calculus. Let us verify that the map indeed preserves the scalar product. We have
\begin{align}
\big (\sqrt{S_\omega} \, \xi_i \otimes |\bm x\rangle, \sqrt{S_\omega} \, \xi_j \otimes |\bm x'\rangle \big ) & =   \big (\xi_i \otimes |\bm x\rangle, S_\omega \xi_j \otimes |\bm x'\rangle \big ) \\ \nonumber
& = \xi_i^T \widehat S_{\bm x, \bm x'}(\omega) \xi_j = S_{\bm x, \bm x'}^{ij}(\omega).
\end{align}
As a consequence,
\begin{equation}
\big (\sqrt{S_\omega} \, \xi_i \otimes |\bm x\rangle, \sqrt{S_\omega} \, \xi_j \otimes |\bm x'\rangle \big ) = \int_{\RM^3}{\rm d}^3\bm r \ \phi_i^\ast(\bm r -\bm x) \phi_j(\bm r - \bm x'),
\end{equation}
for all $i,j = 1,\ldots, N$, as desired.

\vspace{0.2cm}

To preserve the covariance of the physical observables w.r.t. the space group $\GM$ and the disordered configurations, it is important to choose the atomic orbitals as such that they span a linear space, which is closed under representations of the $O(3)$ group. Thus, all $\phi_n$ are assumed to transform under rotations as
\begin{equation}
( \phi_1 , \ldots, \phi_N )\circ \mathfrak r^{-1} = (\phi_1, \ldots, \phi_N) \widehat \Dd(\mathfrak r), \quad \mathfrak r \in O(3) \subset \EM,
\end{equation}
where $\{\widehat \Dd(\mathfrak r), \ \mathfrak r \in O(3)\}$ is a family of $N\times N$ matrices supplying a $N$-dimensional unitary representation of the rotation group that is not necessarily irreducible (see  Section~\ref{Sec:NumSpec} for details). Then, if $\mathfrak g =(\mathfrak p|\mathfrak a) \in \GM$, the overlap matrix satisfies the relation
\begin{equation}\label{Eq:Cov0}
\widehat \Dd (\mathfrak p)^\dagger\widehat S_{\mathfrak g\bm x, \mathfrak g\bm x'}(\tau_{\mathfrak g}\omega) \widehat \Dd(\mathfrak p)= \widehat S_{\bm x, \bm x'}(\omega),
\end{equation}
for any $\bm x$ and $\bm x'$ in $\Ll_\omega$. This is an important relation for which we provide the derivation below. Indeed, both $\mathfrak g \bm x$ and $\mathfrak g \bm x'$ belong to $\mathfrak g \Ll_\omega = \Ll_{\tau_{\mathfrak g}\omega}$ and
\begin{align}\label{Eq:Ex1}
S_{\mathfrak g\bm x, \mathfrak g \bm x'}^{ij}(\tau_\mathfrak g \omega)  
& = \int_{\RM^3}{\rm d}^3\bm r \ \phi_i^\ast(\bm r -\mathfrak g \bm x) \phi_j(\bm r - \mathfrak g \bm x') \\ \nonumber 
& = \int_{\RM^3}{\rm d}^3(\mathfrak g \bm r) \ \phi_i^\ast(\mathfrak g \bm r -\mathfrak g\bm x) \phi_j(\mathfrak g \bm r - \mathfrak g\bm x').
\end{align}
Since $\mathfrak g \bm r - \mathfrak g \bm x = \mathfrak p(\bm r - \bm x)$ and ${\rm d}^3 (\mathfrak g\bm r)={\rm d}^3 \bm r$, we can continue as
\begin{align}\label{Eq:Ex2}
S_{\mathfrak g\bm x, \mathfrak g \bm x'}^{ij}(\tau_{\mathfrak g}\omega) 
& = \int_{\RM^3}{\rm d}^3 \bm r \ \phi_i^\ast\big (\mathfrak p(\bm  r -\bm x\big ) \phi_j\big (\mathfrak p (\bm r - \bm x'\big ) \\ \nonumber
& = \int_{\RM^3}{\rm d}^3 \bm r \ \Dd(\mathfrak p^{-1})^\ast_{ki}\phi_k^\ast(\bm r -\bm x)  \phi_s(\bm r - \bm x') \Dd(\mathfrak p^{-1})_{sj} \\ \nonumber
& = \Dd(\mathfrak p)_{ik}\int_{\RM^3}{\rm d} \bm r \ \phi_k^\ast(\bm r -\bm x) \phi_s(\bm r - \bm x') \Dd(\mathfrak p)^\ast_{js},
\end{align}
and \eqref{Eq:Cov0} follows.

\vspace{0.2cm}

For any $\omega \in \Omega$, we define a Hilbert space isomorphism between $\CM^N \otimes \ell^2(\Ll_\omega)$ and $\CM^N \otimes \ell^2(\Ll_{\tau_{\mathfrak g}\omega})$ as
\begin{equation}
T_{\mathfrak g}\big (\xi \otimes |\bm x\rangle \big ) = \widehat \Dd(\mathfrak p) \xi \otimes |\mathfrak g \bm x \rangle, \quad \mathfrak g = (\mathfrak p|\mathfrak a) \in \GM, \quad \bm x \in \Ll_\omega.
\end{equation}
Note that we use here  the same notation as in \eqref{Eq:EAction} because these two maps can be easily differentiated from the context. Now, by examining the rule of multiplication in \eqref{Eq:GMultiplication} for the space group, it is immediate to see that $T$ respect this binary operation: $T_{\mathfrak g} T_{\mathfrak g'}= T_{\mathfrak g \mathfrak g'}$. Furthermore, it follows directly from \eqref{Eq:Cov0} and the definition \eqref{Eq:OverlapOp} that
\begin{equation}\label{Eq:Cov100}
T_{\mathfrak g} S_\omega T_{\mathfrak g}^\dagger = S_{\tau_{\mathfrak g} \omega}, \quad \mathfrak g \in \GM.
\end{equation}
Indeed, if $\mathfrak g = (\mathfrak p|\mathfrak a)$, then
\begin{align}
T_{\mathfrak g} S_\omega T_{\mathfrak g}^\dagger & = \sum_{\bm x,\bm x' \in \Ll_\omega} \widehat D(\mathfrak p) \widehat S_{\bm x, \bm x'}(\omega) \widehat D(\mathfrak p)^\dagger \otimes |\mathfrak g \bm x \rangle \langle \mathfrak g \bm x' | \\ \nonumber 
& = \sum_{\bm x,\bm x' \in \Ll_{\tau_{\mathfrak g}\omega}}  \widehat D(\mathfrak p^{-1})^\dagger \widehat S_{\mathfrak g^{-1} \bm x, \mathfrak g^{-1}  \bm x'}(\omega) \widehat D(\mathfrak p^{-1}) \otimes | \bm x \rangle \langle  \bm x' |.
\end{align}
Furthermore, from \eqref{Eq:Cov0},
\begin{equation}
\widehat D(\mathfrak p^{-1})^\dagger \widehat S_{\mathfrak g^{-1} \bm x, \mathfrak g^{-1}  \bm x'}(\omega) \widehat D(\mathfrak p^{-1}) = \widehat S_{\bm x,\bm x'}(\tau_{\mathfrak g}\omega),
\end{equation}
hence \eqref{Eq:Cov100} follows.

\vspace{0.2cm}

We conclude this section with the observation that the map $U_\omega$, defined in \eqref{Eq:Map}, satisfies the covariant relation $U_{\tau_{\mathfrak g} \omega} T_{\mathfrak g} = T_{\mathfrak g} U_\omega $.  Indeed, for $\bm x \in \Ll_\omega$,
\begin{align}
\phi_j(\mathfrak g^{-1} \bm r - \bm x)  = \phi_j\big (\mathfrak p^{-1}( \bm r - \mathfrak g \bm x) \big ) 
 = \sum_{k=1}^N \phi_k(\bm r - \mathfrak g\bm x) \Dd(\mathfrak p)_{kj},
\end{align}
while
\begin{align}
T_{\mathfrak g} \sqrt{S_{\omega}} \, \xi_j \otimes |\bm x\rangle  = \sqrt{S_{\tau_{\mathfrak g}\omega}} \, \widehat \Dd(\mathfrak p) \xi_j \otimes |\mathfrak g \bm x\rangle 
 = \sum_{k=1}^N \sqrt{S_{\tau_{\mathfrak g}\omega}}  \, \xi_k \otimes |\mathfrak g \bm x\rangle \Dd(\mathfrak p)_{kj}.
\end{align}
Then, by applying rule \eqref{Eq:Map} on each terms of the two sums, one can convince oneself that we have the following correspondence
\begin{equation}
\phi_j(\mathfrak g^{-1} \bm r - \bm x) \mapsto T_{\mathfrak g} \sqrt{S_{\omega}} \, \xi_j \otimes |\bm x\rangle, \quad \forall \ \bm x \in \Ll,
\end{equation}
under the $U$ map.

\subsection{Canonical tight-binding form of the observables}
\label{Sec:CanonicalOp}

\vspace{0.2cm}

Let $A$ be an operator defined over $L^2(\RM^3)$. Our goal here is to investigate how to define a canonical approximation as an operator $A_\omega$ over the effective Hilbert space $\CM^N \otimes \ell^2(\Ll_\omega)$. The natural requirement is the matching of all the available matrix elements under the $U_\omega$ map \eqref{Eq:Map}, i.e. 
\begin{equation}\label{Eq:CanonicalR}
\langle \phi_n(\cdot -\bm x)|A |\phi_m(\cdot - \bm x') \rangle = \langle n,\bm x|\sqrt{S_\omega} \, A_\omega \sqrt{S_\omega}| m,\bm x'\rangle,
\end{equation}
for all $n,m=\overline{1,N}$ and $\bm x,\bm x' \in \Ll_\omega$. For convenience, above and throughout, we use the notation $|n,\bm x\rangle$ for $\xi_n \otimes |\bm x \rangle$. Henceforth, let $\widehat A_{\bm x,\bm x'}(\omega)$ be the matrix with the entries
\begin{equation}
A_{\bm x,\bm x'}^{n,m}(\omega) = \int_{\RM^3} {\rm d}^3 \bm r \, \phi_n^\ast(\bm r - \bm x) (A\phi_m)(\bm r - \bm x'),
\end{equation}
which is just the explicit form of the coefficients appearing in the left side of \eqref{Eq:CanonicalR}. We form first the operator
\begin{equation}\label{Eq:TildeA}
\widetilde A_\omega = \sum_{\bm x,\bm x' \in \Ll_\omega} \widehat A_{\bm x,\bm x'}(\omega) \otimes |\bm x\rangle \langle \bm x' |,
\end{equation}
over $\CM^N \otimes \ell^2(\Ll_\omega)$. Then, the solution to \eqref{Eq:CanonicalR} is supplied by
\begin{equation}\label{Eq:CanonicalOp}
A_\omega = S_\omega^{-\frac{1}{2}} \ \widetilde A_\omega \ S_\omega^{-\frac{1}{2}},
\end{equation}
as it readily follows from a direct calculation. We call \eqref{Eq:CanonicalOp} the canonical tight-binding operator associated to the operator $A$ that is defined in the continuum KS theory. Note that under this correspondence, the identity operator is sent to the identity operator.

\vspace{0.2cm}

Now assume that the continuum observable depends on $\omega$ in a covariant fashion. In such a case, we can repeat the calculations leading to \eqref{Eq:Cov0} to prove
\begin{equation}\label{Eq:Cov1}
\widehat \Dd (\mathfrak p)^\dagger\widehat A_{\mathfrak g\bm x, \mathfrak g\bm x'}(\tau_{\mathfrak g}\omega) \widehat \Dd(\mathfrak p)= \widehat A_{\bm x, \bm x'}(\omega).
\end{equation}
This automatically implies that $\widetilde A_\omega$ is a covariant operator under the space group transformations and, since $A_\omega$ in \eqref{Eq:CanonicalOp} is a product of covariant operators, $A_\omega$ is also a covariant operator: 
\begin{equation}
T_{\mathfrak g} A_\omega T_{\mathfrak g}^\dagger = A_{\tau_{\mathfrak g} \omega}.
\end{equation}
Below, we apply this standard procedure to several observables of interest.

\vspace{0.2cm}

As we learned in section~\ref{Sec:ContTh}, the continuum KS-Hamiltonian is a covariant observable. Furthermore, among the standard outputs of AIMD simulations are the matrix elements
\begin{equation}\label{Eq:KSMatElem}
W_{\bm x, \bm x'}^{ij}(\omega) = 
\int_{\RM^3} {\rm d}^3 \bm r \ \phi_i^\ast(\bm r - \bm x) (H^\omega_{\rm KS}\phi_j)(\bm r - \bm x').
\end{equation}
This is precisely the data one needs to define the tight-binding Hamiltonian. Following the above procedure, we define first the operator 
\begin{equation}
\widetilde H_\omega = \sum_{\bm x,\bm x' \in \Ll_\omega} \widehat W_{\bm x, \bm x'}(\omega) \otimes |\bm x \rangle \langle \bm x' |,
\end{equation}
which then supplies the tight-binding expression of the KS-Hamiltonian
\begin{equation}
H_\omega = S_\omega^{-\frac{1}{2}} \, \widetilde H_\omega \, S_\omega^{-\frac{1}{2}}, \quad T_{\mathfrak g} H_\omega T_{\mathfrak g}^\dagger = H_{\tau_{\mathfrak g}\omega}, \quad \mathfrak g \in \GM.
\end{equation}

\vspace{0.2cm}

We now focus on the position operator $\bm X$. At the continuum level of the theory, the matrix elements of the position operator are
\begin{equation}\label{Eq:RCoeff}
\bm R_{\bm x, \bm x'}^{ij}(\omega) = \int_{\RM^3}{\rm d}^3\bm r \ \phi_i^\ast(\bm r -\bm x) \, \bm r \,  \phi_j(\bm r - \bm x').
\end{equation}
Note that these matrix elements depend too on the disordered configuration. They, however, satisfy a different covariant relation
\begin{equation}\label{Eq:Convariant5}
\widehat \Dd(\mathfrak g)^\dagger \widehat{\bm R}_{\mathfrak g\bm x, \mathfrak g \bm x'}(\tau_{\mathfrak g}\omega) \widehat \Dd(\mathfrak g) = \mathfrak p \widehat {\bm R}_{\bm x, \bm x'}(\omega) + \mathfrak a \, \widehat S_{\bm x, \bm x'}(\omega),
\end{equation}
for all $\mathfrak g=(\mathfrak p|\mathfrak a)\in \GM$. The above relation follows from an exercise similar to that below \eqref{Eq:Cov0}. Before going any further, let us explain the notation. Note that $\widehat{\bm R}_{\bm x, \bm x'}(\omega)$ is actually a 3-component column vector with matrices as entries. Then, $\mathfrak p$ in front of it, which is an ordinary $3 \times 3$ matrix, acts naturally on this 3-component vector. Furthermore, $\mathfrak a$ in the second term is viewed as an ordinary 3-component vector from $\RM^3$ such that $\mathfrak a \, \widehat S_{\bm x, \bm x'}(\omega)$ becomes a 3-component vector with matrix entries. Now, as before, we define an operator on $\CM^N \otimes \ell^2(\Ll_\omega)$
\begin{equation}
\widetilde{\bm R}_\omega = \sum_{\bm x,\bm x' \in \Ll} \widehat{\bm R}_{\bm x, \bm x'}(\omega) \otimes |\bm x \rangle \langle \bm x' |,
\end{equation}
which satisfies the covariance relation
\begin{equation}\label{Eq:Covariant13}
T_{\mathfrak g} \widetilde{\bm R}_\omega T_{\mathfrak g}^\dagger = \mathfrak p^{-1} \widetilde{\bm R}_{\tau_{\mathfrak g}\omega} + (\mathfrak p^{-1}\mathfrak a) \, S_{\tau_{\mathfrak g}\omega},
\end{equation} 
as it follows directly from \eqref{Eq:Convariant5}. Then
\begin{equation}
\bm X_\omega = S_\omega^{-\frac{1}{2}} \, \widetilde{\bm R}_\omega \, S_\omega^{-\frac{1}{2}}
\end{equation} 
maps the position operator from $L^2(\RM^3)$ to $\CM^N \otimes \ell^2(\Ll_\omega)$. Furthermore, the mapped position operator satisfies the covariance relation
\begin{equation}\label{Eq:CovRelX}
T_{\mathfrak g} \bm X_\omega T_{\mathfrak g}^\dagger = \mathfrak p^{-1} \bm X_{\tau_{\mathfrak g}\omega} + (\mathfrak p^{-1}\mathfrak a) \, I, \quad \mathfrak g=(\mathfrak p|\mathfrak a)\in \GM,
\end{equation}
where $I$ is the identity operator. The above relation follows directly from \eqref{Eq:Covariant13}. 

\vspace{0.2cm}

Note that, although $\bm X_\omega$ is not entirely a covariant operator, the commutator $[\bm X_\omega,A_\omega]$ is covariant whenever $A_\omega$ is, i.e. 
\begin{equation}\label{Eq:CovRelXC}
T_{\mathfrak g} [\bm X_\omega,A_\omega] T_{\mathfrak g}^\dagger = \mathfrak p^{-1} [\bm X_{\tau_{\mathfrak g}\omega},A_{\tau_{\mathfrak g}\omega}], \quad \mathfrak g=(\mathfrak p|\mathfrak a)\in \GM.
\end{equation}
This will become relevant when we will analyze the charge current operator.

\subsection{The trace per volume}
\label{Sec:TrV}

\vspace{0.2cm}

Over the Hilbert space $L^2(\RM^3)$, the trace per volume of a bounded operator $A$ with continuous kernel $\langle \bm r|A|\bm r'\rangle$ is defined as
\begin{equation}
{\rm Tr}_V\{A\} = \lim_{V \rightarrow \RM^3} \tfrac{1}{V} \int_{V} {\rm d} \bm r \ \langle \bm r|A|\bm r\rangle,
\end{equation}
where, for consistency with the space group, we require that the limit be taken over finite volumes $V$ that are invariant under the point group action. Our goal here is to supply its canonical translation over the effective Hilbert space $\CM^N \otimes \ell^{2}(\Ll_\omega)$. For this, let $A_\omega$ be the tight-binding operator associated to $A$. We claim that
\begin{equation}\label{Eq:CanonicalTr}
{\rm Tr}_V\{A_\omega\} = \frac{1}{V_0}\lim_{V \rightarrow \infty} \frac{1}{|\Ll_\omega \cap V|}\sum_{\bm x \in \Ll \cap V} \sum_{n=1}^N \langle n,\bm x | A_\omega |n,\bm x\rangle
\end{equation} 
supplies the canonical expression. Above, $V_0$ is the volume per Si atom, which is just half of that of the primitive cell, and $|\cdot|$ denotes the cardinal of a set.

\vspace{0.2cm}

Indeed, let us note that $\int_{V} {\rm d} \bm r \, \langle \bm r|A|\bm r\rangle$ coincides with the trace of $A$, when $A$ is restricted over $L^2(V)$. This trace can be alternatively computed as $\sum_i \langle \psi_i|A|\psi_i \rangle$, with $\{\psi_i\}$ being an arbitrary orthonormal basis of $L^2(V)$. But, up to errors that are irrelevant in the thermodynamic limit and when $N$ is large, the finite-volume trace can be computed using the partial basis $\{U_\omega^\ast |n,\bm x\rangle\}_{\bm x \in \Ll_\omega \cap V}^{n=\overline{1,N}}$. As a consequence, if $N_a$ is the total number of atoms in $V$, then
\begin{equation}
{\rm Tr}_V\{A\} = \frac{1}{V_0} \lim_{N_a \rightarrow \infty} \frac{1}{N_a} \sum_{n=1}^N \sum_{\bm x \in \Ll \cap V} \langle n,\bm x|U_\omega A U_\omega^\ast|n,\bm x\rangle,
\end{equation}
which coincides with \eqref{Eq:CanonicalTr}.

\vspace{0.2cm}

The trace per volume, which is defined in \eqref{Eq:CanonicalTr}, is a genuine trace over the algebra of operators we encounter in this work. For example, it displays the standard property ${\rm Tr}_V\{A_\omega B_\omega\} = {\rm Tr}_V\{B_\omega A_\omega\}$. An extremely important property of ${\rm Tr}_V$ is the self-averaging when evaluated on covariant operators, {\it i.e.} those operators satisfying the relations
\begin{equation}
T_{\mathfrak g} A_\omega T_{\mathfrak g}^\dagger = A_{\tau_{\mathfrak g}\omega}.
\end{equation}
Indeed, using the invariance of the trace under conjugations, we have
\begin{equation}
{\rm Tr}_V\{A_\omega\} = \frac{1}{|H|}\sum_{\mathfrak g \in H \subset \GM} {\rm Tr}_V\{T_{\mathfrak g} A_\omega T_{\mathfrak g}^\dagger\} = \frac{1}{|H|}\sum_{\mathfrak g \in H \subset \GM} {\rm Tr}_V\{A_{\tau_{\mathfrak g}\omega}\},
\end{equation} 
where $H$ is a finite subset of $\GM$ invariant to the point group. Since $\tau$ acts ergodically over $\Omega$, in the limit $H \rightarrow \GM$, Birkhoff's ergodic theorem assures us that the last term coincides with the ensemble average \cite{Bir}. Hence,
\begin{equation}
{\rm Tr}_V\{A_\omega\} = \int_\Omega {\rm d}\PM (\omega) \ {\rm Tr}_V\{A_\omega\}.
\end{equation}
Let us point out that intensive thermodynamic variable as measured in laboratories, such as the transport coefficients, are all computed as traces per volumes of covariant observables. The aforementioned self-averaging property assures us that these macroscopic variables do not fluctuate from one disordered configuration to another, as long as the corresponding physical observables are covariant. This is the main reason why we pay special attention to the covariant properties of the physical observables in our theory.

\section{Transport Coefficients}

With the mappings from the previous section, we can formulate the theory of quantum charge transport directly on the Hilbert space $\CM^N \otimes \ell^2(\Ll_\omega)$. The goal of this section is to supply the key elements of this theory and to formulate the Kubo-formula for the conductivity tensor. 

\subsection{Kinetic theory of quantum transport}

\vspace{0.2cm}

The purpose of this section is to review the theory of charge transport in the presence of dissipation, as developed by Schulz-Baldes and Bellissard \cite{BellissardJMP1994,SBaldesJSP1998,SBaldesRMP1998}.

\vspace{0.2cm}

Let us recall that the physical observable corresponding to the 3-component vector of the electron charge current density is
\begin{equation}
\bm J_\omega =-\frac{e}{\imath \hbar} [\bm X_\omega,H_\omega],
\end{equation}
where $e=1.6 \times 10^{-19}$~C is the charge of the electron. Based on the last remark in section \ref{Sec:CanonicalOp}, $\bm J_\omega$ is a covariant operator, i.e. 
\begin{equation}
T_{\mathfrak g} \bm J_\omega T_{\mathfrak g} = \mathfrak p^{-1} \bm J_{\tau_{\mathfrak g}\omega}.
\end{equation}
 Under the action of an externally applied electric field $\bm E$, the measured current-density is
\begin{equation}\label{Eq:Current1}
\bm j_{\bm E} =\lim_{T\rightarrow \infty} \frac{1}{T}\int_0^T {\rm d}t \  {\rm Tr}_V\big \{ \bm J_\omega \rho_\omega(t) \big \},
\end{equation}
where $\rho_\omega(t)$ is the time-evolved density matrix. The time evolution is w.r.t. the time-dependent Hamiltonian
\begin{equation}
H(t) = H_\omega + e \bm E \cdot \bm X_\omega + V_\omega(t),
\end{equation}
which incorporates the externally applied electric field $\bm E$, as well as dissipation via the scattering potential
\begin{equation}
V_\omega(t)=\sum_{j \in \ZM} \delta(t-t_j) W_\omega,
\end{equation}
with $W_\omega$ assumed to be covariant. The collision times $\eta=\{t_j\}_{j \in \ZM}$ are generated via a Poisson process with fixed collision time-scale $\tau_c$. Such processes are known to be self-averaging, hence the time and the space averages in \eqref{Eq:Current1} do not depend on the particular realization of the collision times, nor on the disordered configuration. In other words, $\bm j_{\bm E}$ defined in \eqref{Eq:Current1} is a genuine macroscopic thermodynamic coefficient.

\vspace{0.2cm}

For the reason state above, one can use in \eqref{Eq:Current1} an effective quantum time evolution, which is averaged over the Poisson processes $\eta$. A computation of this average can be found in \cite{ProdanAMRX2013}. It takes the form
\begin{equation}\label{Eq:TimeProp}
U_{\rm eff}(t)AU_{\rm eff}(t)^\ast = e^{-\frac{t}{\hbar}(\Gamma_\omega + {\rm L}_{\bm E,\omega})}[A],  
\end{equation}
where $\Gamma$ is the collision super-operator, acting on the physical observables as
\begin{equation}
\Gamma_\omega[A] = \tfrac{\hbar}{\tau_c} (A-e^{\frac{\imath}{\hbar} W_\omega}A \, e^{-\frac{\imath}{\hbar} W_\omega}),
\end{equation} 
and ${\rm L}_{\bm E,\omega}$ is the super-operator
\begin{equation}
{\rm L}_{\bm E,\omega} \, [A] = \imath [H_\omega, A] -  e \bm E \cdot \imath [\bm X_\omega, A].
\end{equation}

\vspace{0.2cm}

The electrons are assumed initially at the thermal equilibrium, hence the initial density matrix takes the form
\begin{equation}
\rho_\omega(t=0) = \Phi_{\rm FD}(H_\omega;T, \mu),
\end{equation}
where $\Phi_{\rm FD}(\epsilon ; T,\mu)$ is the Fermi-Dirac distribution at temperature $T$ and $\mu$ the chemical potential. The density matrix is evolved via the time propagator \eqref{Eq:TimeProp}, hence
\begin{equation}
\rho_\omega(t) = U(t) \, \Phi_{\rm FD}(H_\omega;T, \mu) \, U(t)^\ast.
\end{equation}  
Since, the two parameters $T$ and $\mu$ are kept fixed, we will omit writing them explicitly.

\subsection{Kubo formula with dissipation}
\label{Sec:KF}

\vspace{0.2cm}

With the inputs supplied in the previous section, \eqref{Eq:Current1} can be evaluated explicitly:
\begin{equation}
\bm J_{\bm E}^\omega = \frac{e^2}{\hbar}{\rm Tr}_V\left \{ \big [\bm X_\omega,H_\omega \big] \big (\Gamma_\omega + {\rm L}_{\bm E,\omega}\big )^{-1}
\big [\bm E \cdot \bm X_\omega ,\Phi_{\rm FD}(H_\omega)\big ] \right \}.
\end{equation}
In the linear regime, this leads to a Kubo-formula with dissipation for the conductivity tensor 
\begin{equation}\label{Eq:KuboFormula}
\sigma^{\alpha \beta}(T,\mu; \omega)= -\pi G_0 {\rm Tr}_V\Big \{ \big [ X^\alpha_\omega,H_\omega \big ] (\Gamma_\omega + {\rm L}_\omega)^{-1}
\big [X^\beta_\omega ,\Phi_{\rm FD}(H_\omega)\big ] \Big \},
\end{equation}
where $\alpha$ and $\beta$ indicate space directions, ${\rm L}_\omega$ is the limit of $L_{\bm E,\omega}$ as $\bm E \rightarrow 0$ and $G_0=\frac{2e^2}{h}=7.74 \times 10^{-5}$~S is the conductance quantum. Note that the super-operator $(\Gamma_\omega + {\rm L}_\omega)^{-1}$ acts on the observable appearing at its right.

\subsection{Self-averaging of the transport coefficients}

\vspace{0.2cm}

We now discuss the self-averaging properties of the conductivity tensor. Using the covariant properties of the operators appearing in the Kubo formula, one finds
\begin{align}
\sigma^{\alpha \beta}(T,\mu; \tau_{\mathfrak g}\omega)= &  -\pi G_0 \, \mathfrak p_{\alpha \alpha'} {\rm Tr}_V \Big \{ T_{\mathfrak g} \big [ X^{\alpha'}_\omega,H_\omega \big ] \\ \nonumber
& \qquad \qquad \qquad \qquad (\Gamma_\omega + {\rm L}_\omega)^{-1}
\big [X^{\beta'}_\omega ,\Phi(H_\omega)\big ] T_{\mathfrak g}^\dagger \Big \} \mathfrak p_{\beta' \beta},
\end{align}
for any $\mathfrak g = (\mathfrak p|\mathfrak a) \in \GM$, where repeated indices are summed over their range. Using the invariance of the trace under conjugation, we find the simple rule of transformation
\begin{equation}\label{Eq:SigmaTransform}
\hat \sigma(T,\mu; \tau_{\mathfrak g} \omega) = \mathfrak p \, \hat \sigma(T,\mu; \omega) \, \mathfrak p^{-1}.
\end{equation}
As such, the conductivity tensor is invariant under the translations $(1|\mathfrak t) \in \Bb$. Our observation in section~\ref{Sec:ThDisorder} that this subgroup of $\GM$ acts ergodically on $\Omega$, become extremely important because it assures us that $\hat \sigma(T,\mu)$ is self-averaging and does not fluctuate from one disordered configuration to another. Indeed, given that $\hat \sigma(T,\mu; \omega)=\sigma(T,\mu; \tau_{(1|\mathfrak t)} \omega)$ for any $(1|\mathfrak t) \in \Bb)$, we can write
\begin{align}
\hat \sigma(T,\mu; \omega) & =\lim_{V \rightarrow \RM^3} \frac{1}{|V\cap \Bb|} \sum_{\mathfrak t \in V} \hat \sigma(T,\mu; \tau_{(1|\mathfrak t)} \omega) \\ \nonumber
 & = \int_\Omega {\rm d} \PM(\omega') \, \hat \sigma(T,\mu; \omega'),
\end{align}
where the last equality follows from Birkhoff's theorem \cite{Bir}. Now, the only way to reconcile the above conclusion and \eqref{Eq:SigmaTransform}, is to admit the invariance of the conductivity tensor under the point group action
\begin{equation}
\mathfrak p^{-1} \, \hat \sigma(T,\mu; \omega) \, \mathfrak p = \hat \sigma(T,\mu; \omega), \quad \forall \ \mathfrak p \in \Pp.
\end{equation} 
The remarkable conclusion is that the invariance w.r.t. the full space group $\GM$ of the non-averaged conductivity tensor is exact even though this symmetry is broken locally by the thermal motion of atoms. We mention that in our numerical calculations, we evaluate the isotropic part of the conductivity tensor
\begin{equation}\label{Eq:DCond}
\sigma(T,\mu; \omega) = \tfrac{1}{3} \sum_{\alpha=1}^3 \sigma^{\alpha\alpha}(T,\mu; \omega),
\end{equation}
which is manifestly invariant under the action of the entire space group. 

\vspace{0.2cm}

Let us stress that the above self-averaging property manifests itself only in the strict thermodynamic limit. For finite samples, there will be fluctuations w.r.t. the thermally disordered configurations. This is because the group of symmetry transformations gets reduced when dealing with finite samples and, as a consequence, $\tau_{\mathfrak g}\omega$ does not explore\footnote{Up to subsets of measure zero} the whole $\Omega$ when $\mathfrak g$ is given all allowed values. For a finite Si crystal of cubic shape, which is built by repeating the unit cell, the rank of the group of symmetries is equal to the number $N_a$ of atoms in the crystal. Given the invariance of $\sigma(\mu,T; \omega)$ w.r.t. these transformations, when we evaluate $\sigma(\mu,T; \omega)$ for one disordered configuration, we in fact evaluate the conductivity for all $\tau_{\mathfrak g}\omega$ configurations. In other words, with just one calculation, we sample $N_a$ points of $\Omega$. Hence, if we repeat the calculation of $\sigma(\mu,T; \omega)$ for a number $N_c$ of different configurations, we effectively sampled $\Omega$ at $N_a \times N_c$ points.  Because of this amplification effect that stems from the invariance of $\sigma(\mu,T; \omega)$ relative to the space symmetries, we expect that a good disorder average can be achieved even with a small number of disordered configurations. This is indeed observed in our simulations.

\vspace{0.2cm}

\subsection{Optimal finite-volume approximations}
\label{Sec:FiniteVolApprox}

\vspace{0.2cm}

There are two fundamental difficulties when attempting to evaluate \eqref{Eq:KuboFormula} on a computer. The first one stems from the incompatibility between the covariant relation \eqref{Eq:CovRelX} for the position operator and the periodic boundary conditions. The second difficulty comes from inverting the super-operator $\Gamma_\omega + {\rm L}_\omega$. Both these issues have been resolved in \cite{ProdanAMRX2013} and then further refined in \cite{ProdanSpringer2017,BourneJPA2018}. In the present context, however, the situation is slightly different because the position operator depends on the disordered configuration. This complication is being addressed below.

\vspace{0.2cm}

We start by computing the matrix elements of the commutator of a continuum observable with the position operator:
\begin{align}
[\bm X,A]_{\bm x, \bm x'}^{m,n}(\omega)  = & \int_{\RM^3} {\rm d}^3 \bm r \, \phi_m^\ast(\bm r - \bm x) ([\bm X,A]\phi_n)(\bm r - \bm x') \\ \nonumber
 = & \int_{\RM^3} {\rm d}^3 \bm r \int_{\RM^3} {\rm d}^3 \bm r' \, (\bm r -\bm r')\phi_m^\ast(\bm r - \bm x) A(\bm r,\bm r')\phi_n(\bm r' - \bm x') \\ \nonumber
 = & (\bm x - \bm x')\int_{\RM^3} {\rm d}^3 \bm r \int_{\RM^3} {\rm d}^3 \bm r' \, \phi_m^\ast(\bm r - \bm x) A(\bm r,\bm r')\phi_n(\bm r' - \bm x') \\ \nonumber
& + \int_{\RM^3} {\rm d}^3 \bm r \int_{\RM^3} {\rm d}^3 \bm r' \, (\bm r -\bm r')\phi_m^\ast(\bm r) A(\bm r,\bm r')\phi_n(\bm r').
\end{align}
We can summarize the above calculation as
\begin{equation}\label{Eq:XComm1}
[\bm X,A]_{\bm x, \bm x'}^{m,n}(\omega) = (\bm x - \bm x') A_{\bm x,\bm x'}^{m,n}(\omega) + [\bm X,A]_{\bm 0, \bm 0}^{m,n}(\omega).
\end{equation}
While the right hand side makes perfect sense for an infinite samples, when the simulation proceeds over a finite crystal with periodic boundary conditions, there is an obvious problem with the first term. In \cite{ProdanSpringer2017,BourneJPA2018}, it was found that the optimal adaptation to the periodic boundary conditions is through the following substitution:
\begin{equation}\label{Eq:Subst}
\bm x - \bm x' \rightarrow \bm x - \bm x' - \left [ \left [\frac{\bm x - \bm x'}{L/2} \right ] \right ] L,
\end{equation}
where $L$ is the size of the periodic super-cell of the simulation and $[[\cdot]]$ denotes the integer part of a real number. The second term in \eqref{Eq:XComm1} is a local term and there is no need for a modification when finite crystals with  periodic boundary conditions are considered. With the proper matrix elements at hand, the finite-volume tight-binding operators corresponding to the commutators with the position operator are derived via the procedure detailed in section~\ref{Sec:CanonicalOp} without any modifications. To alert the reader about the substitution \eqref{Eq:Subst}, we write the modified commutators of these tight-binding operators as $\lfloor X_\omega ,A_\omega \rfloor$. 

\vspace{0.2cm}

We now focus on the super-operator $\Gamma_\omega + {\rm L}_\omega$. We will only consider here the so called relaxation time approximation where the dissipation super-operator is proportional with identity: $\Gamma_\omega = \Gamma_0 \, {\rm id}$, with $\Gamma_0$ a positive number. Now we recall that ${\rm L}_\omega$ acts on operators $A_\omega$ over $\CM^N \otimes \ell^2(\Ll_\omega)$ via ${\rm L}_\omega [A_\omega] = \imath [H_\omega,A_\omega]$. Observe that, if
\begin{equation}
\big(\epsilon_a^\omega, \ \psi_a^\omega\big )_{a =1,\ldots, N|\Ll_\omega|}
\end{equation}
 is an eigen-system for $H_\omega$, then
\begin{equation}
{\rm L}_\omega \big [|\psi_a^\omega\rangle \langle \psi_b^\omega| \big ] = \imath (\epsilon_a^\omega - \epsilon_b^\omega) \, |\psi_a^\omega\rangle \langle \psi_b^\omega| .
\end{equation}
In other words,
\begin{equation}
\Big (\epsilon_a^{\omega} - \epsilon_b^{\omega}, \ |\psi_a^{\omega}\rangle \langle \psi_b^{\omega}|\Big )_{a,b = 1,\ldots,N|\Ll_\omega|}
\end{equation} 
is an eigen-system for ${\rm L}_\omega$. This observation together with the fact that any operator can be decomposed as
\begin{equation}
A_\omega = \sum_{a,b} \langle \psi_a^\omega | A_\omega |\psi_b^\omega \rangle \, |\psi_a^\omega \rangle \langle \psi_b^\omega |
\end{equation} 
provide a straightforward way to invert the super-operator:
\begin{equation}
(\Gamma_0 \, {\rm id} + {\rm L}_\omega)^{-1}[A_\omega] = \sum_{a,b} \frac{\langle \psi_a^\omega | A_\omega |\psi_b^\omega \rangle}{\Gamma_0 + \imath (\epsilon_a^\omega - \epsilon_b^\omega)} |\psi_a^\omega \rangle \langle \psi_b^\omega |. 
\end{equation}

\vspace{0.2cm}

Finally, we can give a direct translation of the Kubo-formula \eqref{Eq:KuboFormula} at finite-volume:
\begin{equation}
\sigma^{\alpha \beta} = \Big \langle \tfrac{-\pi G_0}{\rm Vol} \sum_{a,b} \frac{ \big \langle \psi_a^\omega\big | \lfloor X^\alpha_\omega,H_\omega\rfloor \big | \psi_b^\omega \big \rangle \big \langle \psi_b^\omega \big | \lfloor X^\beta_\omega, \Phi_{\rm FD}(H_\omega)\rfloor \big | \psi_a^\omega \big \rangle }{\Gamma_0 + \imath(\epsilon_a^\omega - \epsilon_b^\omega)} \Big \rangle_\omega.
\end{equation}
This expression is useful when the matrix elements of the Fermi operator are available. Since this quantity is not among the standard outputs of AIMD simulations, we process this expression one step further as in \cite{ProdanAOP2016}:
\begin{align}
\sigma^{\alpha \beta} = \Big \langle \tfrac{-\pi G_0}{\rm Vol} \sum_{a,b} & \frac{\Phi_{\rm FD}(\epsilon_a^\omega)-\Phi_{\rm FD}(\epsilon_b^\omega)}{\epsilon_a^\omega - \epsilon_b^\omega}  \\ \nonumber
&  \quad \times \, \frac{ \big \langle \psi_a^\omega\big | \lfloor X^\alpha_\omega,H_\omega \rfloor \big | \psi_b^\omega \big \rangle \big \langle \psi_b^\omega \big | \lfloor X^\beta_\omega, H_\omega \rfloor \big | \psi_a^\omega \big \rangle }{\Gamma_0 + \imath(\epsilon_a^\omega - \epsilon_b^\omega)} \Big \rangle_\omega.
\end{align}
This is the expression we coded as a post-processing subroutine to the AIMD simulations. The inputs for this expression are the matrix elements of the Kohn-Sham Hamiltonians \eqref{Eq:KSMatElem} and the overlap coefficients \eqref{Eq:Overlap}, as well as the xyz-coordinates of the atoms.

\section{Numerical Implementation}

In this section, we first present a novel electronic structure method that is only scaling quadratically with system size, thus facilitating second-generation Car-Parrinello AIMD simulations of even longer length and time scales than previously thought feasible \cite{CPMD2, TDK2014}. More importantly, this approach permits to efficiently compute the exact finite-temperature density matrix $\rho_{\omega}$ of a given Kohn-Sham Hamiltonian $H_{\omega}$  ``on-the-fly'' during the AIMD. Thereafter, the computational details of our simulatios are described in detail. 

\subsection{Field-Theory-based Eigenvalue Solver}

\vspace{0.2cm}

Following Alavi and coworkers \cite{frenkel, alavi}, we begin with the (Helmholtz) free energy functional 
\begin{equation}
  \mathcal{F} = \Xi + \mu N_{e} + V_{dc}, \label{FreeEnerFunc}
\end{equation}
where $N_{e} = 2\Nn$ is the number of electrons and $\Xi$ the grand-canonical potential (GCP) for noninteracting fermions. The latter reads as 
\begin{equation}\label{GCPbasic}
  \Xi = -\tfrac{2}{\beta}\, \textup{ln~det} \left( 1 + e^{\beta\left(\mu S_{\omega} - H_{\omega} \right)} \right) = - \tfrac{2}{\beta}\, \text{Tr}~\textup{ln} \left( 1 + e^{\beta\left(\mu S_{\omega} - H_{\omega} \right)} \right), 
\end{equation}
with given by $\beta^{-1} = k T$ ($k=$ Boltzmann constant). Yet, in the low-temperature limit 
\begin{equation}
  \lim_{\beta \rightarrow \infty} {\Xi} = 2 \sum_{a=1}^{\Nn} {\epsilon^\omega_{a}} - \mu N_{e}, \label{GCPlowTempLim}
\end{equation}
the so-called band-structure energy, which is given by the sum of the lowest $\Nn$ doubly occupied eigenvalues $\epsilon^\omega_{a}$ of $H_{\omega}$, can be recovered and 
\begin{equation}
\lim_{\beta \rightarrow \infty} {\mathcal{F}} = 2 \sum_{a=1}^{\Nn} \epsilon^\omega_{a} + V_{dc}
\end{equation}
holds. Therein, $V_{dc}$ accounts for double counting terms, as well as for the nuclear Coulomb interaction. 

In the present case of fully self-consistent KS-DFT calculations
\begin{eqnarray}
  V_{dc}[n_{\omega}(\bm{r})] &=& - \tfrac{1}{2} \int{ \, d\bm{r} \int{ \, d\bm{r}' \, \frac{n_{\omega}(\bm{r}) n_{\omega}(\bm{r}') }{|\bm{r}-\bm{r}'|} } } \nonumber \\
  &-& \int{\, d\bm{r} \, n_{\omega}(\bm{r}) \, \frac{\delta \Xi_{\text{XC}}[n_{\omega}(\bm{r})]}{\delta n_{\omega}(\bm{r})}} + \Xi_{\text{XC}}[n_{\omega}(\bm{r})] + E_{II}, \label{dcDFT}
\end{eqnarray}
where the first term on the right hand side is the double counting correction of the Hartree energy, while $\Xi_{\text{XC}}[n_{\omega}(\bm{r})]$ is the finite-temperature XC grand-canonical functional and $E_{II}$ the nuclear Coulomb interaction. Except for the latter term, Eq.~(\ref{dcDFT}) accounts for the difference between $\Xi$ and the GCP for the interacting spin-$\frac{1}{2}$ Fermi gas, i.e. 
\begin{eqnarray}
  \Xi_{int}[n_{\omega}(\bm{r})] &=& -\tfrac{2}{\beta} \ln \det \left( 1 + e^{\beta\left(\mu S_{\omega} - H_{\rm KS}^\omega \right)} \right) \nonumber \\
  &-& \tfrac{1}{2} \int{ \, d\bm{r} \int{ \, d\bm{r}' \, \frac{n_{\omega}(\bm{r}) n_{\omega}(\bm{r}') }{|\bm{r}-\bm{r}'|} } } \nonumber \\
  &-& \int{\, d\bm{r} \, n_{\omega}(\bm{r}) \, \frac{\delta \Xi_{\text{XC}}[n_{\omega}(\bm{r})]}{\delta n_{\omega}(\bm{r})}} + \Xi_{\text{XC}}[n_{\omega}(\bm{r})].
\end{eqnarray}
As before, in the low-temperature limit $\Xi_{int}[n_{\omega}(\bm{r})] + \mu N_{e}$ equals to the band-structure energy, whereas $\Xi_{\text{XC}}[n_{\omega}(\bm{r})]$ corresponds to the familiar XC energy, so that in this limit $\mathcal{F} = \Xi + \mu N_{e} + V_{dc} = \Xi_{int}[n_{\omega}(\bm{r})] + \mu N_{e} + E_{II}$ is equivalent to the Harris-Foulkes functional \cite{harris, foulkes}.
Equally than the latter, $\mathcal{F}$ is explicitly defined for any $n_{\omega}(\bm{r})$ and obeys exactly the same stationary point as the finite-temperature functional of Mermin \cite{mermin}. 

Whereas it is well known how to calculate $V_{dc}$ with linear-scaling computational effort, the computation of all occupied orbitals by diagonalization requires $\mathcal{O}(N^{3})$ operations. Due to the fact that the band-structure term can be equivalently expressed in terms of $\rho_{\omega}$, the total energy can be written as 
\begin{equation}
  E_{\rm KS}[n_{\omega}(\bm{r})] = 2 \sum_{a=1}^{\Nn} \epsilon^\omega_{i} + V_{dc} = \text{Tr} [ \rho_{\omega} H_{\rm KS}^\omega ] + V_{dc}[n_{\omega}(\bm{r})]. \label{TotEnerMat}
\end{equation}
As a consequence, the cubic-scaling diagonalization of $H_{\rm KS}^\omega$ can be bypassed by directly calculating $\rho_{\omega}$ rather than all $\varepsilon_{i}$'s. 

In order to make further progress, let us now factorize the operator of Eq.~(\ref{GCPbasic}) into $P$ terms. Given that $P$ is even, which we shall assume in the following, Krajewski and Parrinello derived the following identity 
\begin{equation}
1 + e^{\beta\left(\mu S_{\omega} - H_{\rm KS}^\omega \right)} = \prod_{l=1}^{P} \left(1 - e^{\frac{i\pi}{2P}\left(2l-1\right)} e^{\frac{\beta}{2P}\left(\mu S_{\omega} - H_{\rm KS}^\omega \right)}\right)  
= \prod_{l=1}^{P/2} {M}_l^* {M}_l, \label{GCPdecomp}
\end{equation}
where the matrices ${M}_l$, with $l = 1, \ldots, P$, are defined as
\begin{equation} 
  {M}_l := {1} - e^{\frac{i\pi}{2P}\left(2l-1\right)} e^{\frac{\beta}{2P}\left(\mu S_{\omega} - H_{\rm KS}^\omega \right)}, \label{MlMat}
\end{equation}
while $^*$ denotes complex conjugation \cite{kraj1}. Analog to numerical path-integral calculations \cite{John}, it is possible to exploit the fact that if $P$ is large enough, so that the effective temperature $\beta/P$ is small, the exponential operator $e^{\frac{\beta}{2P}\left(\mu S_{\omega} - H_{\rm KS}^\omega \right)}$ can be approximated by a Trotter decomposition or simply by a high-temperature expansion, i.e. 
\begin{equation}
  {M}_{l} = {1}-e^{\frac{i\pi}{2P}\left(2l-1\right)} \left( {1} + \tfrac{\beta}{2P}(\mu S_{\omega} - H_{\rm KS}^\omega) \right) + \mathcal{O} \left( \frac{1}{P^2} \right). \label{approxMlMat}
\end{equation}
However, as we will see, here no such approximation is required, which is in contrast to the original approach \cite{kraj1}. 
In any case, the GCP can be rewritten as 
\begin{eqnarray}
\Xi &=& -\tfrac{2}{\beta} \textup{ln} \prod_{l=1}^{P/2} \textup{det}\, ({M}_l^* {M}_l) = \tfrac{4}{\beta} \sum_{l=1}^{P/2} \textup{ln} \left( \textup{det}\, ({M}_l^* {M}_l) \right)^{-\frac{1}{2}}. 
\label{GCP}
\end{eqnarray}

As is customary in lattice gauge theory \cite[p.~17]{mont}, where the minus sign problem is avoided by sampling a positive definite distribution, the inverse square root of the determinant can be expressed as an integral over a complex field $\bm{\phi}_{l}$, which has the same dimension $M$ as the full Hilbert space, i.e. 
\begin{eqnarray}
  \det \left({M}_{l}^{*} {M}_{l} \right)^{-1/2} = \frac{1}{(2\pi)^{\frac{M}{2}}} \int d\phi_l \, e^{-\frac{1}{2} \phi_l^* {M}_l^* {M}_l \phi_l}. \label{InvDet}
\end{eqnarray}
Inserting Eq.~(\ref{InvDet}) into Eq.~(\ref{GCP}) we end up with the following field-theoretic expression for the GCP: 
\begin{eqnarray}
  \Xi &=& \tfrac{4}{\beta} \sum_{l=1}^{P/2} \textup{ln}\, \left[ \frac{1}{(2\pi)^{\frac{M}{2}}} \int d\phi_l \, e^{-\frac{1}{2} \phi_l^* {M}_l^* {M}_l \phi_l} \right] \nonumber \\
  &=& \tfrac{4}{\beta} \sum_{l=1}^{P/2} \textup{ln}\, \int d\phi_l \, e^{-\frac{1}{2} \phi_l^* {M}_l^* {M}_l \phi_l} + const., \label{GCPft}
\end{eqnarray}
where $\phi_l$ are appropriate vectors.

All physical relevant observables can then be determined as functional derivatives of the GCP w.r.t. an appropriately chosen external parameter. For example, $N_{e} = - \partial \Xi / \partial \mu$ and $\lim_{\beta \rightarrow \infty}{\Xi} + \mu N_{e} = 2 \sum_{a=1}^{\Nn} \epsilon^\omega_{i}$, so that 
\begin{equation}
  E = \lim_{\beta \rightarrow \infty}{\mathcal{F}} =  2 \sum_{a=1}^{\Nn} \epsilon^\omega_{a} + V_{dc} = \frac{\partial (\beta \Xi)}{\partial \beta} - \mu \frac{\partial \Xi}{\partial \mu} + V_{dc}. \label{GCPtotEner}
\end{equation}
Since the functional derivative of the constant in Eq.~(\ref{GCPft}) is identical to zero, all physical interesting quantities can be computed via 
\begin{subequations}
\begin{eqnarray}
\frac{\partial \Xi}{\partial \lambda} 
&=& -\tfrac{2}{\beta} \sum_{l=1}^{P/2}  \frac{\int d\phi_l \, \sum\limits_{i,j = 1}^d (\phi_l)_i^* \left( \frac{\partial ({M}_l^* {M}_l)}{\partial \lambda} \right)_{ij} (\phi_l)_j \, e^{ - \frac{1}{2} \phi_l^* {M}_l^* {M}_l \phi_l }}{\int d\phi_l \, e^{ -\frac{1}{2} \phi_l^* {M}_l^* {M}_l \phi_l}} \\
&=& -\tfrac{2}{\beta} \sum_{l=1}^{P/2}  \sum_{i,j = 1}^d \left( \frac{\partial ({M}_l^* {M}_l)}{\partial \lambda} \right)_{ij} ({M}_l^* {M}_l)^{-1}_{ij} \\
&=& -\tfrac{2}{\beta} \sum_{l=1}^{P/2} \text{Tr} \left[ ({M}_l^* {M}_l)^{-1} \frac{\partial ({M}_l^* {M}_l)}{\partial \lambda} \right] = -\tfrac{2}{\beta} \sum_{l=1}^{P} \text{Tr} \left[ {M}_l^{-1} \frac{\partial {M}_l}{\partial \lambda} \right]. \qquad \quad \label{GCPfuncDeriv} 
\end{eqnarray}
\end{subequations}
Thereby, the left-hand side of Eq.~(\ref{GCPfuncDeriv}) holds because of Montvay and M\"unster \cite[p. 18]{mont}, 
whereas the right-hand side is due to the fact that beside being positive definite ${M}_l^* {M}_l$ is also symmetric. 


Comparing Eq.~(\ref{TotEnerMat}) with Eq.~(\ref{GCPlowTempLim}), it is easy to see that the GCP and hence all physical significant observables can be written as the trace of a product consisting of the Fermi matrix ${\rho}_{\omega}$. 
Specifically, $\Xi = \text{Tr}[{\rho}_{\omega} H_{\rm KS}^\omega] - \mu N_{e}$, but because at the same time $N_{e} = \text{Tr}[{\rho}_{\omega} S_{\omega}]$ holds, the former can be simplified to 
\begin{equation}
  \Xi = \text{Tr} [{\rho}_{\omega} (H_{\rm KS}^\omega - \mu S_{\omega})], 
\end{equation}
where 
$S_{\omega} = - \partial H_{\rm KS}^\omega / \partial \mu$ and ${\rho}_{\omega} = {\partial \Xi} / {\partial H_{\rm KS}^\omega}$. 
As a consequence, the GCP and all its functional derivatives can be reduced to evaluate ${\rho}_{\omega}$ based on Eq.~(\ref{GCPfuncDeriv}) with $\lambda = H_{ij}$. 
Using the identity 
\begin{eqnarray}
  \frac{\partial {M}_{l}}{\partial H_{ij}} &=& - \tfrac{1}{2P} \left\{ ({M}_{l} - {1}) \beta + \beta ({M}_{l} - {1}) \right\}, 
\end{eqnarray}
for this particular case, Eq.~(\ref{GCPfuncDeriv}) eventually equals to 
\begin{equation}
{\rho}_{\omega} = \frac{\partial \Xi}{\partial H_{\rm KS}^\omega} = \tfrac{4}{P} \sum_{l=1}^{P/2} \left({1} - \bigl({M}_l^* {M}_l \bigr)^{-1}\right) = \tfrac{2}{P} \sum_{l=1}^{P} \left({1} - {M}_l^{-1}\right). \label{DensityMatrix}
\end{equation}
In other words, the origin of the method is the notion that the density matrix, the square of the wavefunction at low temperature and the Maxwell-Boltzmann distribution at high temperature, can be decomposed into a sum of ${M}_l^{-1}$ matrices, each at higher effective temperature $\beta/P$ and hence always sparser than ${\rho}_{\omega}$.  Yet, contrary to the original approach \cite{kraj1}, neither a Trotter decomposition nor a high-temperature expansion for Eq.~(\ref{MlMat}) has been used, so far everything is exact for any $P$.

In particular, unlike Eq.~(\ref{GCP}), the determination of $\Omega = \partial (\beta \Omega) / \partial \beta$ does no longer involve the calculation of the inverse square root of a determinant, but just the inverse of ${M}_{l}$, which 
is not only very sparse, since it obeys the same sparsity pattern as $H_{\rm KS}^\omega$, but is furthermore also always better conditioned as the latter. Hence, all ${M}_{l}^{-1}$ matrices are substantially sparser than ${\rho}_{\omega}$ and thus can be efficiently computed \cite{ceri1, ceri2}.
In fact, for quasi one-dimensional systems, ${M}_{l}$ is tridiagonal that permits for an exact linear-scaling calculation of its inverse using a recursive scheme \cite{Godfrin1991}. For all other dimensions $D$, ${M}_{l}$ can be sought of being block-tridiagonal, where the dimensionality of each block is $d=N^{1-(1/D)}$, eventually leading to a computational effort, which scales like $Nd^2=N^{3-2/d}$. Since this is only marginally better than the initial $\mathcal{O}(N^{3})$ scaling for a general matrix inversion (or diagonalization), we compute ${M}_{l}^{-1}$ by solving the $N_e$ sets of linear equations ${M}_{l} {\Phi}_{j}^{l} = {\psi}_{j}$, where $\{ {\psi}_{j} \}$ is a complete set of basis functions \cite{Richters}. Using a preconditioned biconjugate gradient method \cite{nr}, the inverse can be exactly computed as ${M}_{l}^{-1} = \sum_{j=1}^{N_{e}}{{\phi}_{j}^{l} {\psi}_{j}^{l}}$ within $\mathcal{O}(N^{2})$ operations. Furthermore, the formal analogy of the decomposition to the Trotter factorization immediately suggests the possibility to apply some of the here presented ideas with benefit to numerical path-integral calculations \cite{John}. The same applies for a related area where these methods are extensively used, namely the lattice gauge theory to quantum chromodynamics \cite{kogut}, whose action is rather similar to the one of Eq.~(\ref{InvDet}).

\begin{figure}[t!]
\center
  \includegraphics[width=\textwidth]{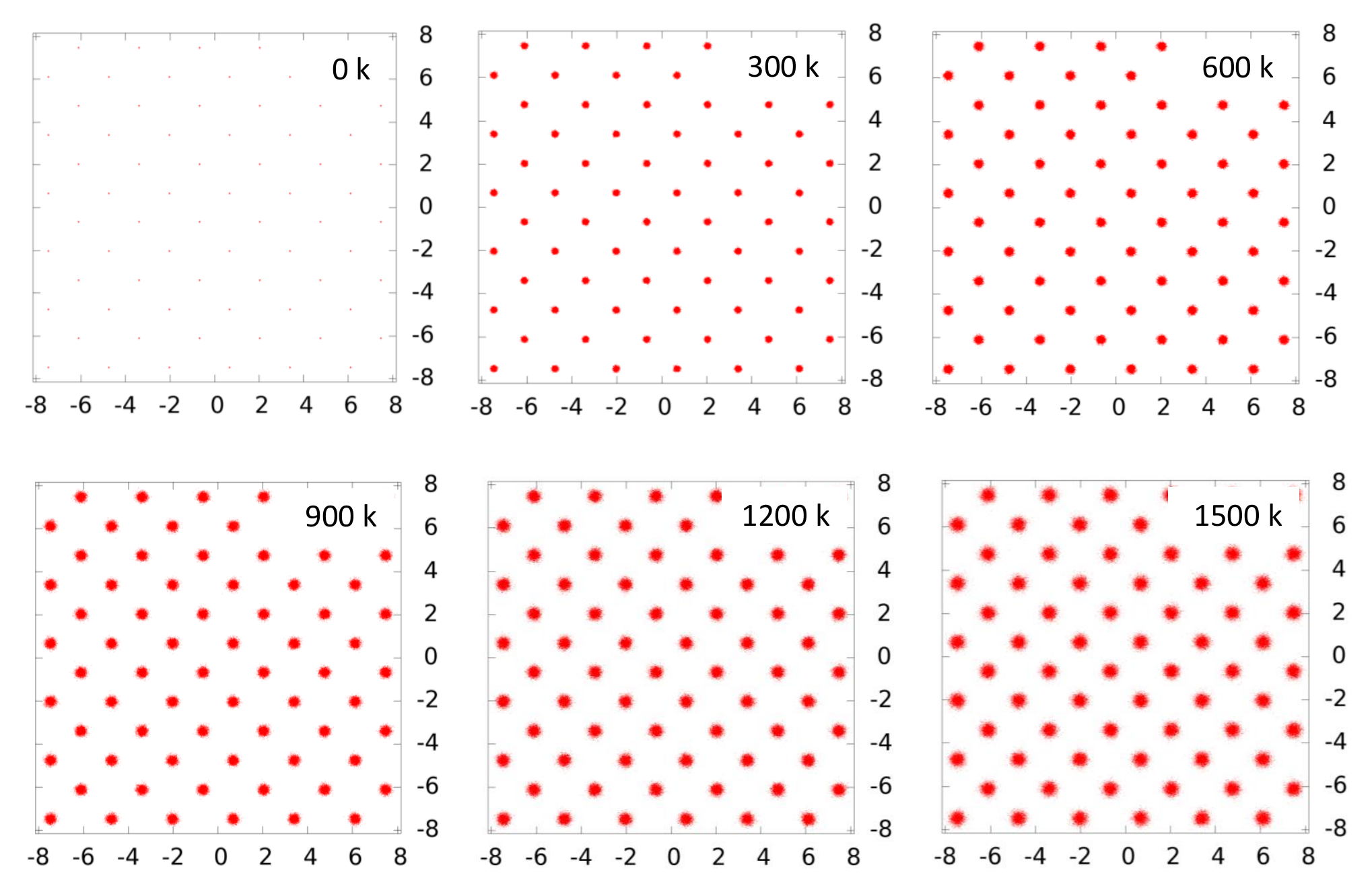}\\
  \caption{\small The orbits of the atoms under the thermal motion at different temperatures. The simulation is for a crystal containing 216 Si atoms and each orbit is sampled at 1001 points. In these renderings, the crystal is viewed from atop of xy-plane. The units of the graphs are Angstroms.}
  \label{Fig:ThermalMotion216}
\end{figure}

\subsection{Computational Details}
\label{Sec:NumSpec}

\vspace{0.2cm}

We now return to our specific simulations. Our models of crystalline silicon consisted of 216 and 1000 Si atoms in a cubic simulation box with periodic boundary conditions. For each system size, five simulations have been conducted, at $T=300$~K, $600$~K, $900$~K, $1200$~K  and $1500$~K, respectively. All of our calculations were performed in the canonical NVT ensemble using the second-generation Car-Parrinello AIMD method of K\"uhne and coworkers \cite{CPMD2,TDK2014}. Throughout, the experimental density of crystalline silicon was assumed, which, at ambient conditions, is semiconducting and four-fold coordinated. In each run, we carefully equilibrated the system for 250~ps before accumulating statistics during additional 1.25~ns, resulting in a total AIMD simulation time of 15~ns. 

\vspace{0.2cm}

All simulations were performed at the DFT level using the mixed Gaussian and plane wave code CP2K/\textit{Quickstep} \cite{Quickstep}. In this approach, the KS orbitals are expanded in contracted Gaussians functions, while the electronic charge density is represented by plane waves \cite{GPW}. 
A density cutoff of 100~Ry was employed for the latter, whereas for the former a dimer-optimized minimal basis set was used \cite{MOLOPT} of $s$- and $p$-type. As such, $N=4$ in \eqref{Eq:LocalHilbert} and the linear space spanned by these wave functions is indeed a representation space for the $O(3)$ group. The unknown exact XC potential is substituted by the LDA \cite{DMC}, whereas the interactions between the valence electrons and the ionic cores are described by separable norm-conserving Goedecker-Teter-Hutter pseudopotentials \cite{GTH,PP}.  For the sake of simplicity, the first Brillouin zone of the super cell is sampled at the $\Gamma$-point only.

\begin{figure}[t!]
\center
  \includegraphics[width=\textwidth]{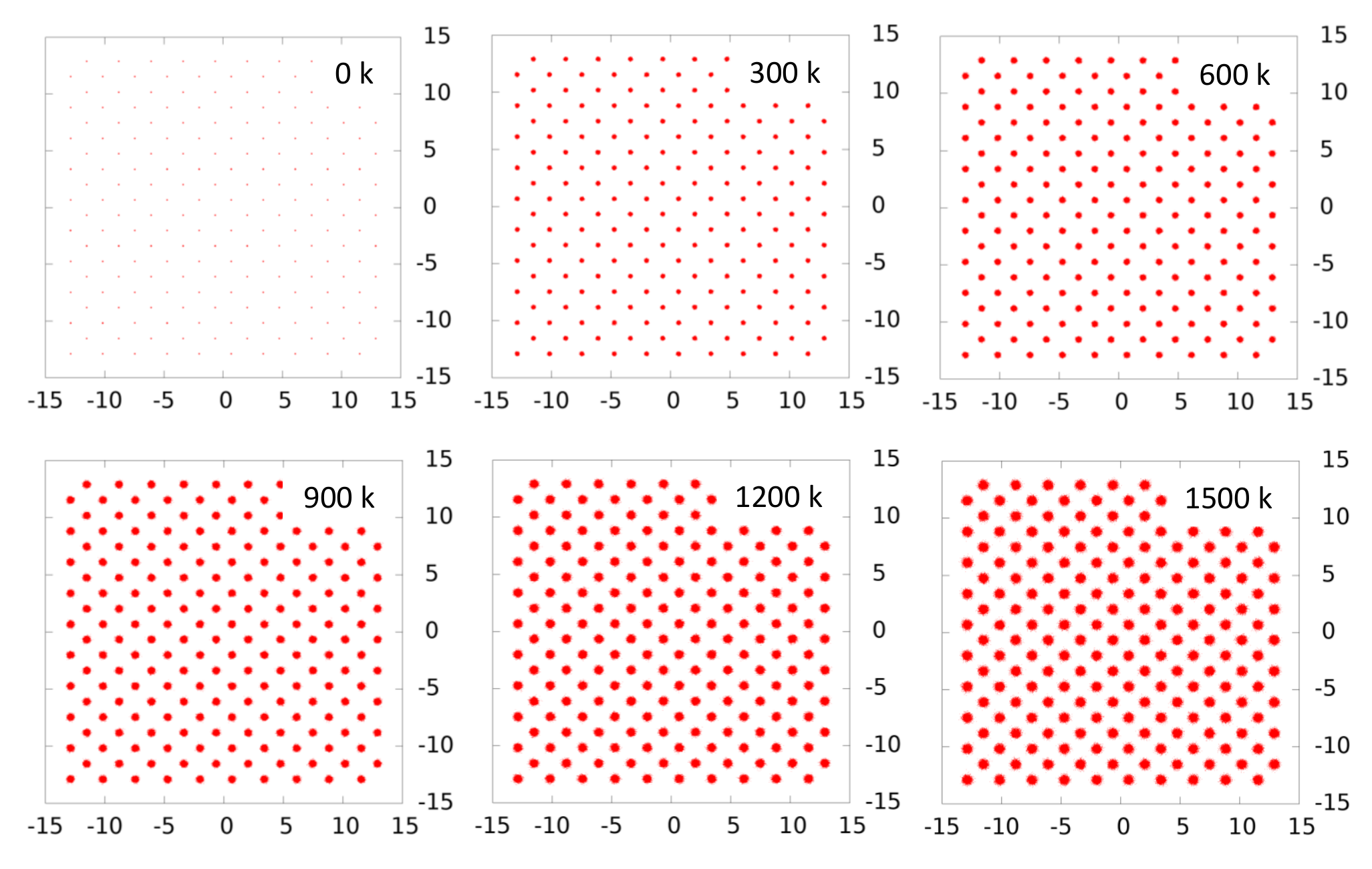}\\
  \caption{\small Same as Fig.~\ref{Fig:ThermalMotion1000} but for a simulation with 1000 Si atoms.}
  \label{Fig:ThermalMotion1000}
\end{figure}

\section{Numerical results}

In this section, we first present and analyze the output of our AIMD simulations and then we report the output of the charge-transport post-processing subroutine detailed in section~\ref{Sec:FiniteVolApprox}. 

\subsection{Spectral analysis}

\vspace{0.2cm}

In Figs.~\ref{Fig:ThermalMotion216} and \ref{Fig:ThermalMotion1000} we report the orbits of the atoms at different temperatures, as simulated with 216 and 1000 Si atoms, respectively. As one can see, the orbits wonder around the equilibrium positions and the data reveals that crystal Si is quite disordered even at the room temperature. Let us point out that many electronic devices operate at 600~K or higher under heavy loads. At these temperatures, the thermal disorder is quite pronounced. As it is well known, in such conditions, some of the wave functions can and will become affected by the phenomenon called Anderson localization \cite{AbrahamPRL1979}. When a wave function becomes Anderson localized, its contribution to the Kubo-formula is zero. Almost as a rule \cite{AizenmanRMP1994}, these localized states occur close to the edges of the energy spectrum and, for 3-dimensional crystals, it is predicted there exist mobility edges in the energy spectrum, one in the conduction and one in valence bands,  beyond which the wave functions remain extended. These mobility edges define the so called mobility gap and the expectation of the charge current operator is zero when one only populates electron states with eigen-energy within this gap. It becomes clear that the activated behavior of the conductivity is determined by the mobility gap and not by the spectral gap. As such, it is extremely important to detect the mobility edges for our crystals. For this, we employ a technique called the level statistics analysis, which has been successfully used in the past for this very purpose \cite{ProdanSpringer2017,ProdanJPA2011}. 

\begin{figure}[t]
\center
  \includegraphics[width=\textwidth]{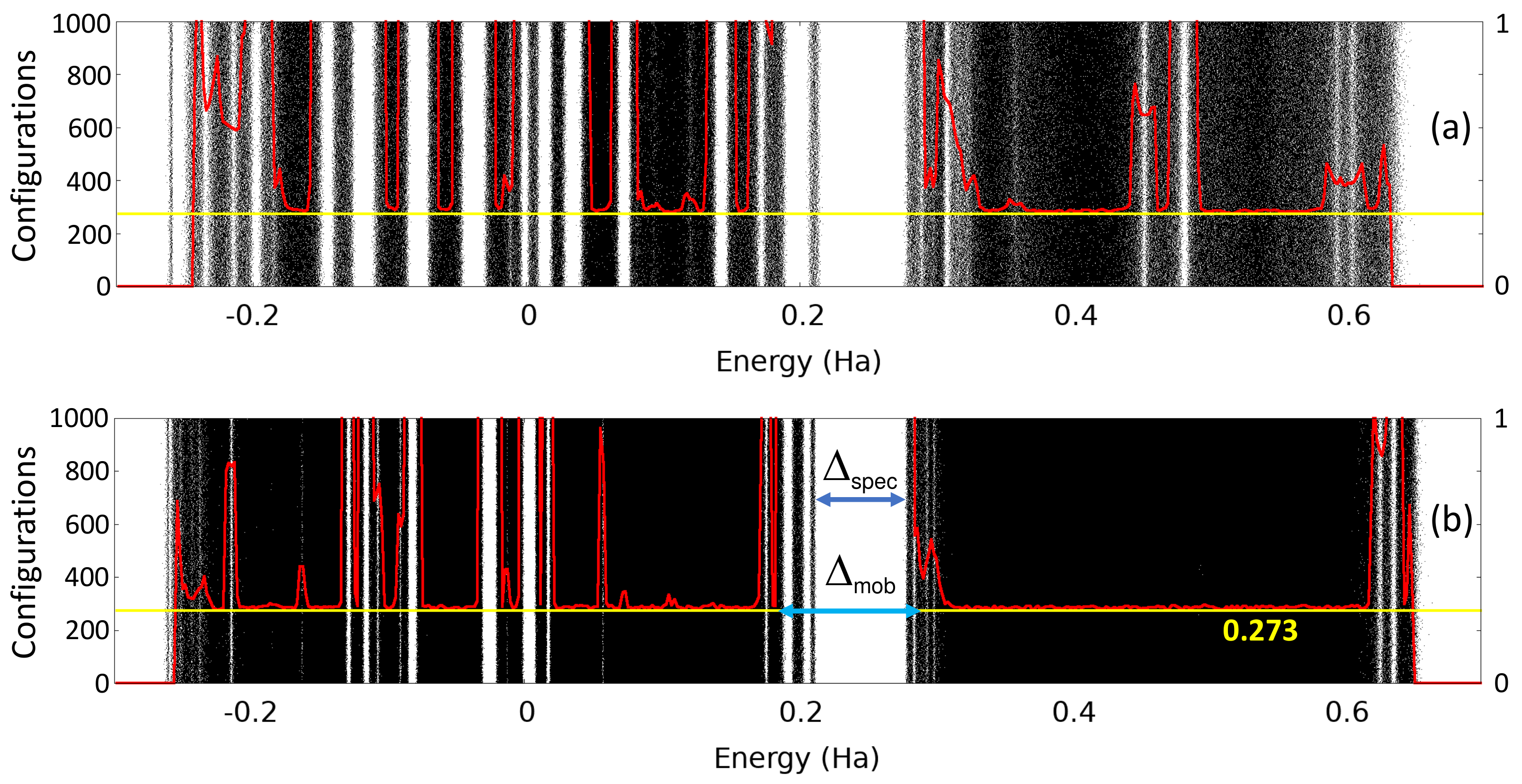}\\
  \caption{\scriptsize Level statistics for a (a) $3 \times 3 \times 3$ unit cells crystal containing 216 atoms and (b) $5 \times 5 \times 5$ unit cells crystal containing 1000 atoms. Both sets of data were generated at $T = 900$~K. The background displays the energy spectra for various thermally disordered configurations. The red curve represents the variance of the level spacings ensembles collected at different energies. The yellow line indicates the variance of the Gaussian orthogonal ensemble. Both the red and yellow curves have their y-axis on the right side. Also shown are the spectral and the mobility gaps, as inferred from the data.}
  \label{Fig:LevelStat}
\end{figure}

\vspace{0.2cm}

We exemplify the process for the temperature $T=900$~K, where the disorder is quite pronounced and the effects described above are more visible. Before we start, we need to examine the spectral characteristics of the Hamiltonians. For this, we have diagonalized the tight-binding KS-Hamiltonian for 1001 selected thermally disordered configurations. The result is a sequence of 1001 discrete sets of eigenvalues $\{\epsilon_a^\omega\}$, which we rendered on a horizontal line for each configuration and then we stuck these lines vertically. The resulting collection of spectra then appears as the fuzzy dots seen in the background of Fig.~\ref{Fig:LevelStat}. We recall that for covariant systems of Hamiltonians in the thermodynamic, the spectrum is in fact non-fluctuating in the sense that, if we pick any energy interval and ask what is the probability (w.r.t. to $\omega$) for at least one eigenvalue to fall within this interval, the answer will either 0 or 100 percent. This is a consequence of the fact that $T_{\mathfrak g} H_\omega T_{\mathfrak g}^\dagger$ have the same spectrum for all $\mathfrak g \in \GM$. In the same time, $T_{\mathfrak g} H_\omega T_{\mathfrak g}^\dagger = H_{\tau_{\mathfrak g}\omega}$ and the orbit $\{\tau_{\mathfrak g}\omega, \, \mathfrak g \in \GM\}$ samples $\Omega$ entirely. The spectrum of a covariant family of Hamiltonians is defined as the intersection of all closed subsets of the real axis that contain all eigenvalues with 100\% probability. Rendering the spectra as in Fig.~\ref{Fig:LevelStat} helps one identify this non-fluctuating spectral set.

\begin{figure}[t!]
\center
  \includegraphics[width=0.8\textwidth]{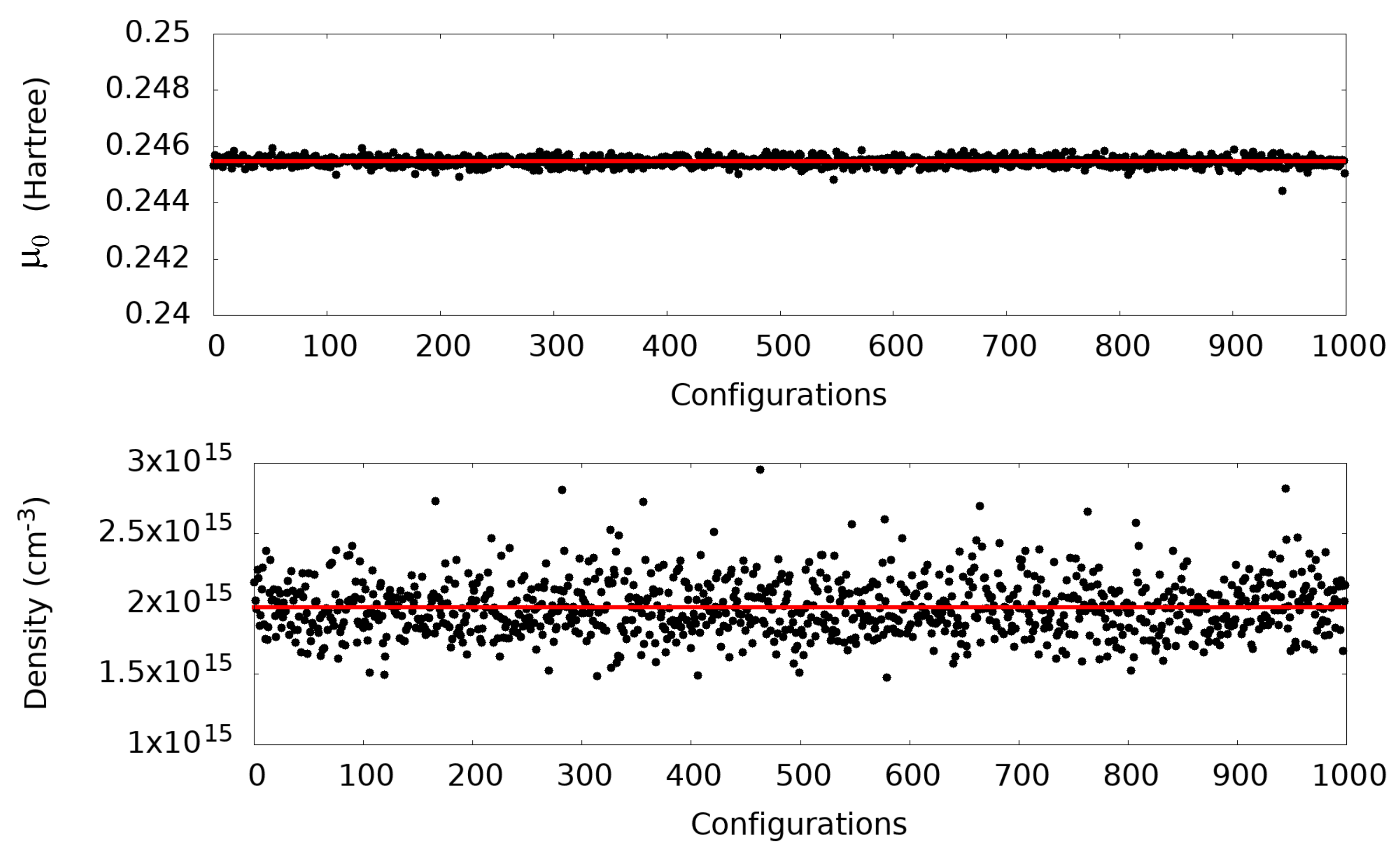}\\
  \caption{\scriptsize (a) Fluctuation (black dots) and average value (red line) of the chemical potential value corresponding to the charge neutrality. (b) Same as (a) for the intrinsic charge carrier density. The data was extracted from the spectra shown in Fig.~\ref{Fig:LevelStat}(b) for a crystal containing 1000 Si atoms at temperature 900~K.}
  \label{Fig:IntrinsicDen}
\end{figure} 

\vspace{0.2cm}

We now examine the spectra more closely. The first issue we want to address is the appearance of the spectral gaps inside the valence and conduction bands. These are artificial features due to relative small size of the system. For a periodic system simulated with periodic boundary conditions ({\it i.e.} at $\Gamma$-point) on a finite super-cell containing many unit cells, this gaps will be explained by the coarse sampling of the Brillouin zone of the unit cell. As one can see in Fig.~\ref{Fig:LevelStat}, many of the gaps disappear when the size of the Si crystal is increased from $3 \times 3 \times 3$ unit cells to $5 \times 5 \times 5$ unit cells. One should also note that the fuzziness in the rendered spectra decreases as the size of the system increases, which is a manifestation of the non-fluctuating character of the spectrum in the thermodynamic limit. 

\vspace{0.2cm}

The second issue is the size of the spectral gap, which in our simulations comes at $1.7$ eV. This is more than twice the value returned by converged KS-DFT simulations and it indicates that the local orbital basis is too coarse. We have verified that, indeed, increasing the local orbital basis converges the spectral gap to the standard KS-DFT value of 0.7 eV. We recall that the experimental value is 1.1700 eV at 4.2K \cite{SpringerHandbook} and that the experimental band gap displays a temperature dependence which has been assessed quite precisely \cite{VarshniPhys1967,ODonnellAPL1991}. Our simulations, however, are performed with the same super-cell regardless of the temperature, hence we cannot relate them to that experimental fact. Our conclusions based on the spectral data reported in Fig.~\ref{Fig:LevelStat} is that the present simulations are not yet precise enough for quantitative predictions. As such, we will focus in the following only at qualitative aspects.

\vspace{0.2cm}

We now turn our focus on the level statics analysis, which was performed in the following way. We picked an arbitrary energy $\epsilon$ and, for each of the 1000 thermally disordered configurations considered in Fig.~\ref{Fig:LevelStat}, we identified the unique eigenvalues $\epsilon_a^\omega$ and $\epsilon_{a+1}^\omega$ that satisfy the constraint $\epsilon_a^\omega < \epsilon < \epsilon_{a+1}^\omega$. Then we computed the level spacings $\Delta \epsilon$=$\epsilon_{a+j+1}^\omega- \epsilon_{a+j}\omega$, letting $j$ take 11 consecutive values between $-5$ and $5$. After repeating the procedure for all 1000 configurations, we generated ensembles of 11,0000 level spacings for each energy $\epsilon$. These level spacings were subsequently normalized by their average.

\vspace{0.2cm}

As done in \cite{ProdanJPA2011}, one can examine the histograms of these ensembles and determine what kind of distributions they manifest. Since the KS-Hamiltonians are real, we expect the outcome to be either a Poisson distribution $P(s)=e^{-s}$ or a Gaussian orthogonal ensemble (GOE), $P_{\mbox{\tiny{GOE}}}=\frac{\pi}{2}s e^{-\frac{\pi}{4}s^2}$. These distributions are expected when the localization length of the wave functions with energy close to $\epsilon$ is smaller/larger than the size of the super-cell, respectively \cite{EfetovBook}. If the super-cell is large enough, one can derive from these distributions the localized or de-localized character of the wave functions. Overlapped over the spectra in Fig.~\ref{Fig:LevelStat} is the variance $\langle s^2 \rangle - \langle s \rangle ^2$ of the level spacing ensembles collected at different energies, as well as the variance value of 0.273 computed from $P_{\mbox{\tiny{GOE}}}$. Since the variance of the Poisson ensemble is 1, we can easily identify from Fig.~\ref{Fig:LevelStat} the character of the wave functions, in particular, the mobility gap. As one can see, it extends well beyond the spectral gap.

\begin{figure}[t!]
\center
  \includegraphics[width=0.7\textwidth]{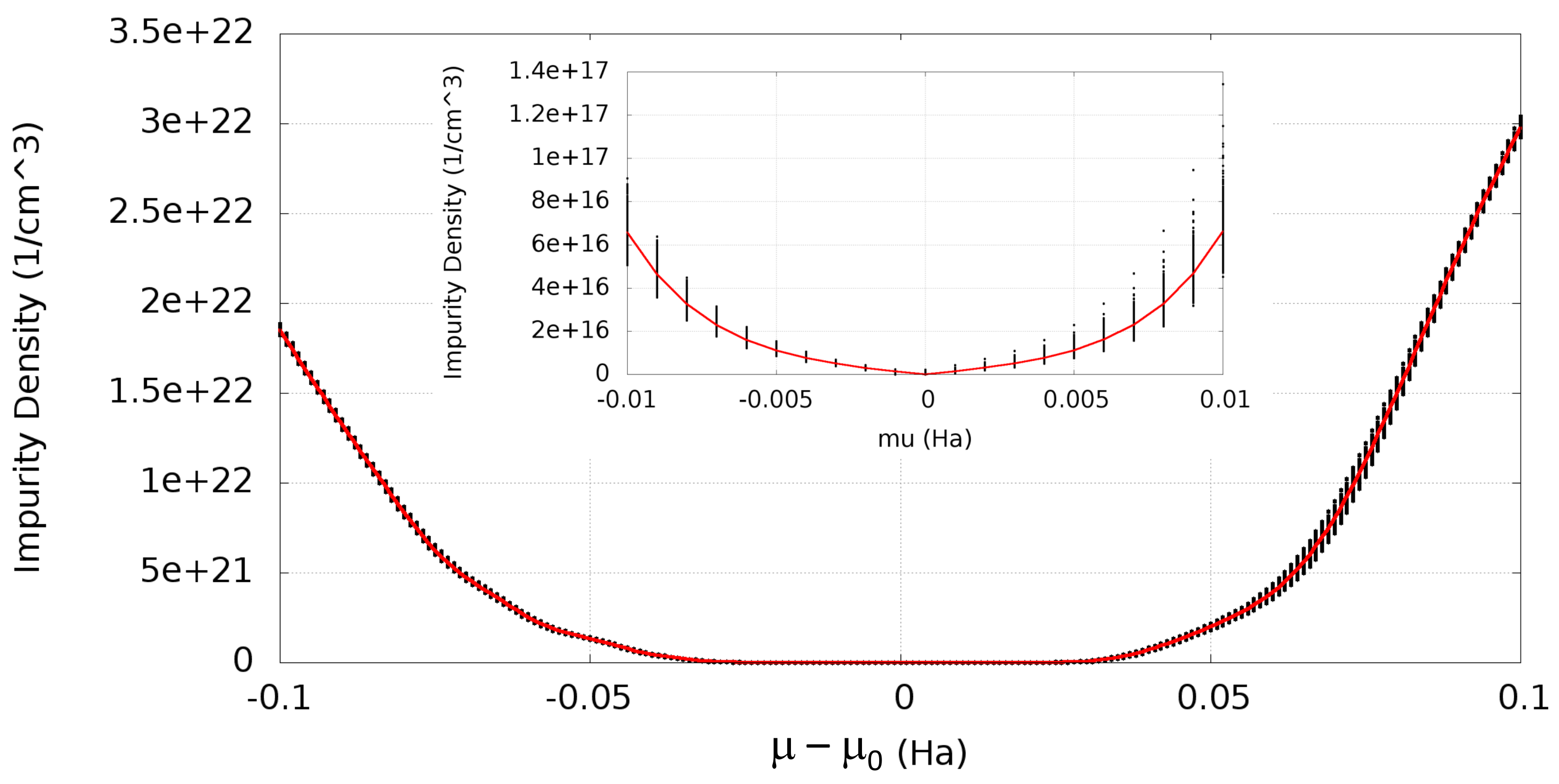}\\
  \caption{\scriptsize Fluctuation (black dots) and average value (red line) of the impurity density as a function of chemical potential. The inset shows a restricted range of the same data. The numerical values were extracted from the spectra shown in Fig.~\ref{Fig:LevelStat}(b) for a crystal containing 1000 Si atoms at temperature 900~K.}
  \label{Fig:ExtrinsicDen}
\end{figure}

\subsection{Charge carrier concentrations} 

\vspace{0.2cm}

\begin{figure}[t!]
\center
  \includegraphics[width=0.9\textwidth]{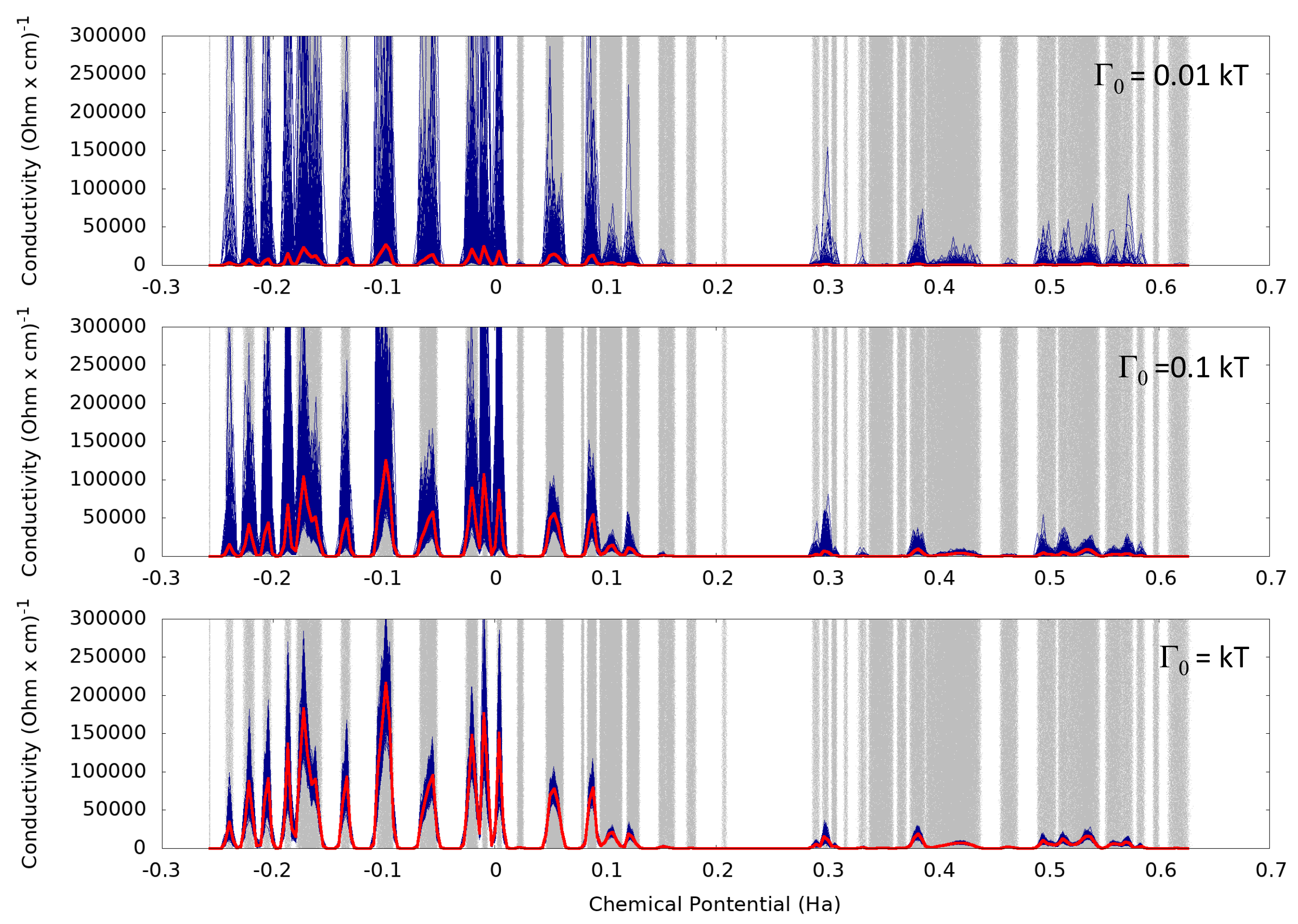}\\
  \caption{\scriptsize Conductivity as a function of chemical potential for 216 Si atoms, $T=300$~K and different values of the dissipation coefficient. Shown in blue are the un-processed output for 1000 thermally disordered configurations. The red curves represent the average values.}
  \label{Fig:Sigma216At300K}
\end{figure}

The charge-neutrality point is defined by the precise value of the chemical potential $\mu_0$ where the charge neutrality of the crystal is achieved. Since in our calculations each ionic core carries $4e$ charge, the charge neutrality condition reads:
\begin{equation}\label{Eq:IntrDen}
\frac{4N_a}{\rm Vol} = \left \langle \frac{2}{\rm Vol}\sum_{a} \frac{1}{1+ \exp(\frac{\epsilon_a^\omega-\mu_0}{kT})} \right \rangle_\omega.
\end{equation}
We want to point out that, for covariant systems, $\mu_0$ and the quantity inside the average brackets in \eqref{Eq:IntrDen} are self-averaging in the thermodynamic limit. However, for our finite-size crystals, these quantities will display fluctuations from one thermally disordered configuration to another and the size of these fluctuations is a good indicator of how close is the simulation to the thermodynamic limit. A rendering of the fluctuations as well as the average value of the chemical potential $\mu_0$ at the neutrality point and 900~K temperature are reported in Fig.~\ref{Fig:IntrinsicDen}(a). The data reveal an extremely low level of fluctuations, characterized by a standard deviation of $0.063$\,\%  around the average value $\mu_0=0.24546$~Ha.

\vspace{0.2cm}

\begin{figure}[t!]
\center
  \includegraphics[width=0.9\textwidth]{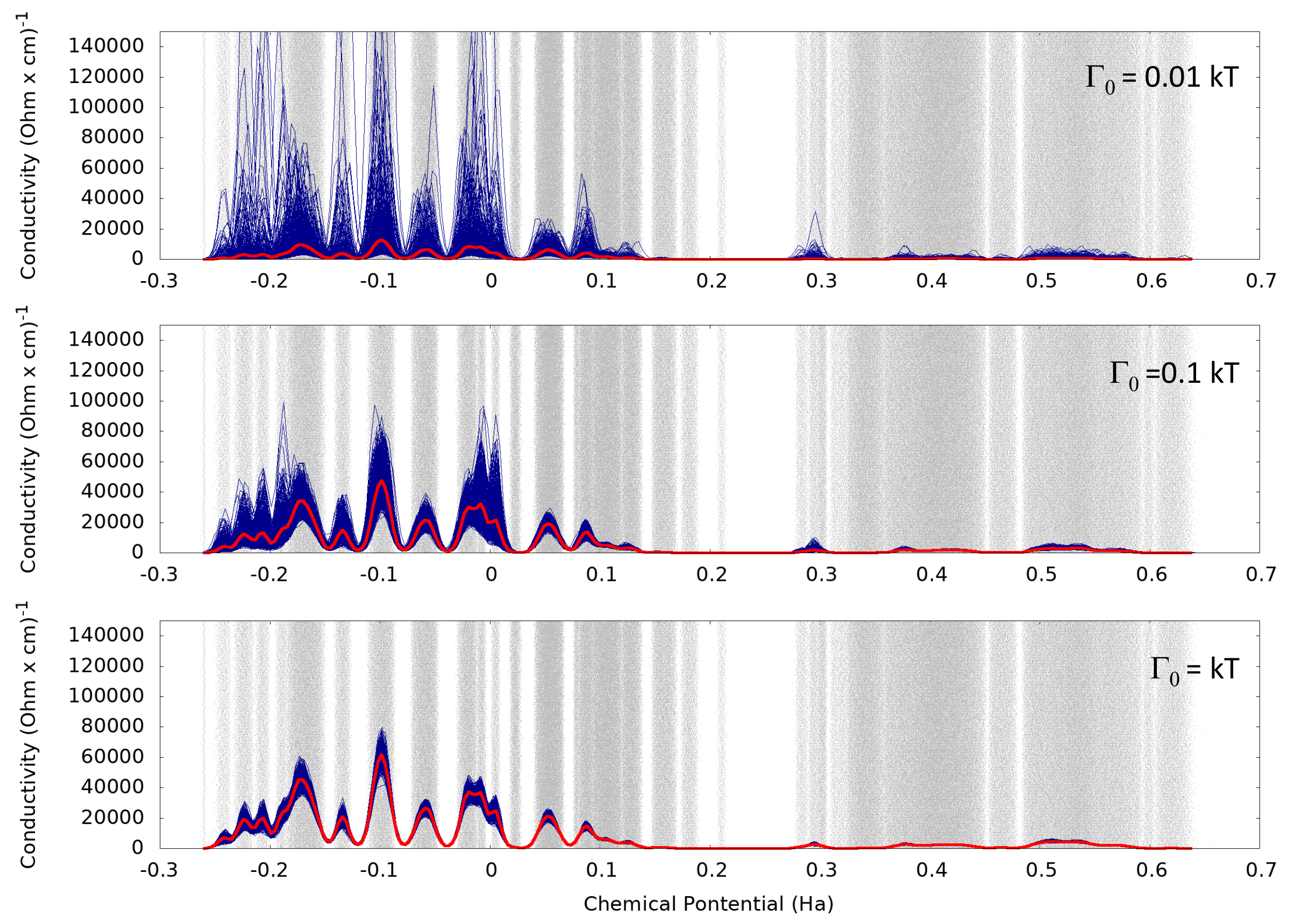}\\
  \caption{\scriptsize Same as Fig.~\ref{Fig:Sigma216At300K} for $T = 900$~K.}
  \label{Fig:Sigma216At900K}
\end{figure}

Mapping the concentration of the conduction electrons and valance holes is crucial for understanding the transport characteristics of crystals. The hole concentration is determined by the depletion of the valence states due to the thermal excitations:
\begin{equation}\label{Eq:HDen}
n_{\rm h}(T,\mu)=\left \langle \frac{2}{{\rm Vol}}\sum_{\epsilon_a^\omega \leq \mu_0} \left [1-\frac{1}{1+ \exp(\frac{\epsilon_a^\omega-\mu}{kT})}\right ] \right \rangle_\omega .
\end{equation}
The concentration of the mobile electrons is determined by the population of the conduction states due to the thermal excitations:
\begin{equation}\label{Eq:EDen}
n_{\rm e}(T,\mu)=\left \langle \frac{2}{{\rm Vol}}\sum_{\epsilon_a^\omega \geq \mu_0} \frac{1}{1+ \exp(\frac{\epsilon_a^\omega-\mu}{kT})} \right \rangle_\omega .
\end{equation}
Note that at the neutrality point, we have the equality: 
\begin{equation}
n_{\rm h}(T,\mu_0) = n_{\rm e}(T,\mu_0).
\end{equation} 
The common value of the two densities is called the intrinsic density ($n_i$) of charge carriers and it is one of the most important characteristic of Si semiconductor. Its experimental value at 300~K has been determined with great precision \cite{SproulJAP1991,MisiakosJAP1993,AltermattJAP2003} to be $n_i=9.7 \div 10.0 \times 10^9$~cm$^{-3}$. Experimental data on the dependence of $n_i$ with the temperature has been summarized in \cite[Fig.~14]{Thurmond1975}, from where we extracted the experimental value of $1.0\times 10^{17}$~cm$^{-3}$ at $T=900$~K. For our simulations, the fluctuations and the average value of $n_i$ are reported in Fig.~\ref{Fig:IntrinsicDen}(b). As one can see, despite of extremely low fluctuations in $\mu_0$, there are substantial fluctuations in the $n_i$ data, which reflect the extreme sensitivity of $n_i$ on the energy spectrum. Quantitatively, the standard deviation in Fig.~\ref{Fig:IntrinsicDen}(b) is 6\,\% and the average value is $n_i = 1.973 \times 10^{15}$. This value is much lower than the experimental value mentioned above, the main reason being the over-estimation of the band gap by our simulations.

\begin{figure}[t!]
\center
  \includegraphics[width=0.9\textwidth]{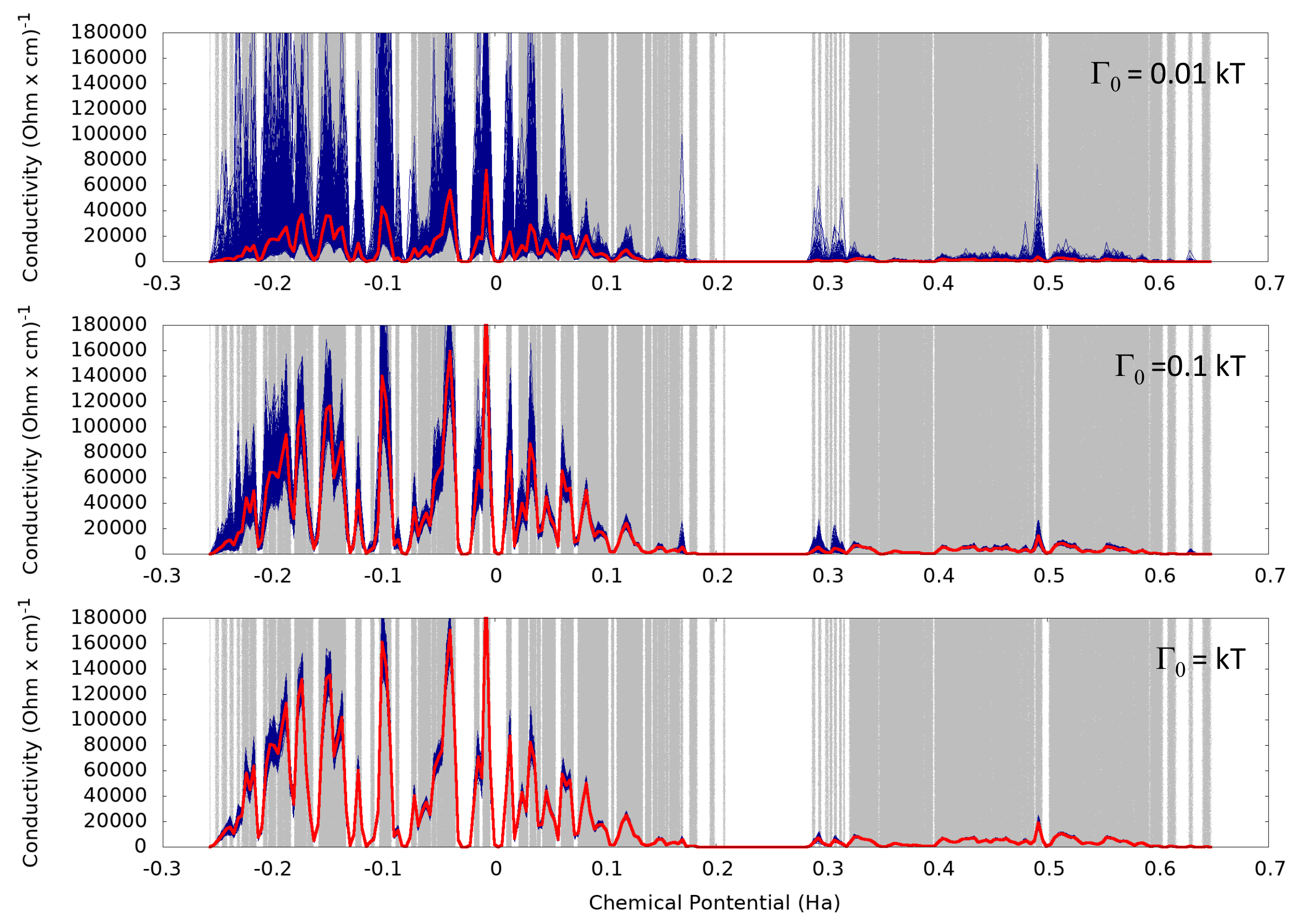}\\
  \caption{\scriptsize Direct conductivity as a function of the chemical potential for 1000 Si atoms, $T=300$~K and $\Gamma_0=0.01kT$ (top), $\Gamma_0=0.1kT$ (middle) and $\Gamma_0=kT$ (bottom). Shown in blue are the un-processed output for 1000 thermally disordered configurations. The red curves represent the average values.}
  \label{Fig:Sigma1000At300K}
\end{figure}

\vspace{0.2cm}

In our study, we will consider not only neutral but also Si crystals that are away from neutrality point by letting the chemical potential $\mu$ be a variable. Experimentally, the variation of the chemical potential can be achieved via gate potentials for thin films or via impurity doping for bulk samples. Either way, such variations lead to changes in the electronic structure of the crystal, which should be recomputed every time the doping level is changed. Since in our simulations we use the same electronic structure, specifically the one computed at the neutrality point, the results we present here are relevant only for lightly doped or weakly gated samples where the changes in the electronic structure are expected to insignificant. To make contact with the experiment, one has to rely on the impurity density value rather than on the chemical potential, because the former is the parameter that can be controlled in laboratory. The impurity density is evaluated from:
\begin{equation}\label{Eq:ExtrinsicDen}
n(\mu,T)= |n_{\rm e}(\mu,T)-n_{\rm h}(\mu,T)|=\left |\left \langle \frac{2}{{\rm Vol}}\sum_{a} \frac{1}{1+ \exp(\frac{\epsilon_a^\omega-\mu}{kT})} \right \rangle_\omega - \frac{4N_a}{\rm Vol} \right |.
\end{equation}
For completeness, we show in Fig.~\ref{Fig:ExtrinsicDen} the relation between the impurity density and chemical potential $\mu$, as derived from the spectra shown in Fig.~\ref{Fig:LevelStat} and \eqref{Eq:ExtrinsicDen}. Let us recall that a light to moderate doping corresponds to the experimental values $n_{\rm exp} < 10^{16}$~cm$^{-3}$, which in terms of the chemical potential means, approximately, that $| \mu-\mu_0 | < 0.01\,{\rm Ha}$.

\vspace{0.2cm} 

Let us end this section by specifying that the extrinsic hole density from Eqs.~\eqref{Eq:HDen} will be used in the next section to generate the hole mobility via the relation $\sigma = e n_{\rm h} \mu_{\rm h}$. The hole mobility $\mu_{\rm h}$ will be mapped as a function of the acceptor concentration \eqref{Eq:ExtrinsicDen}.

\begin{figure}
\center
  \includegraphics[width=0.9\textwidth]{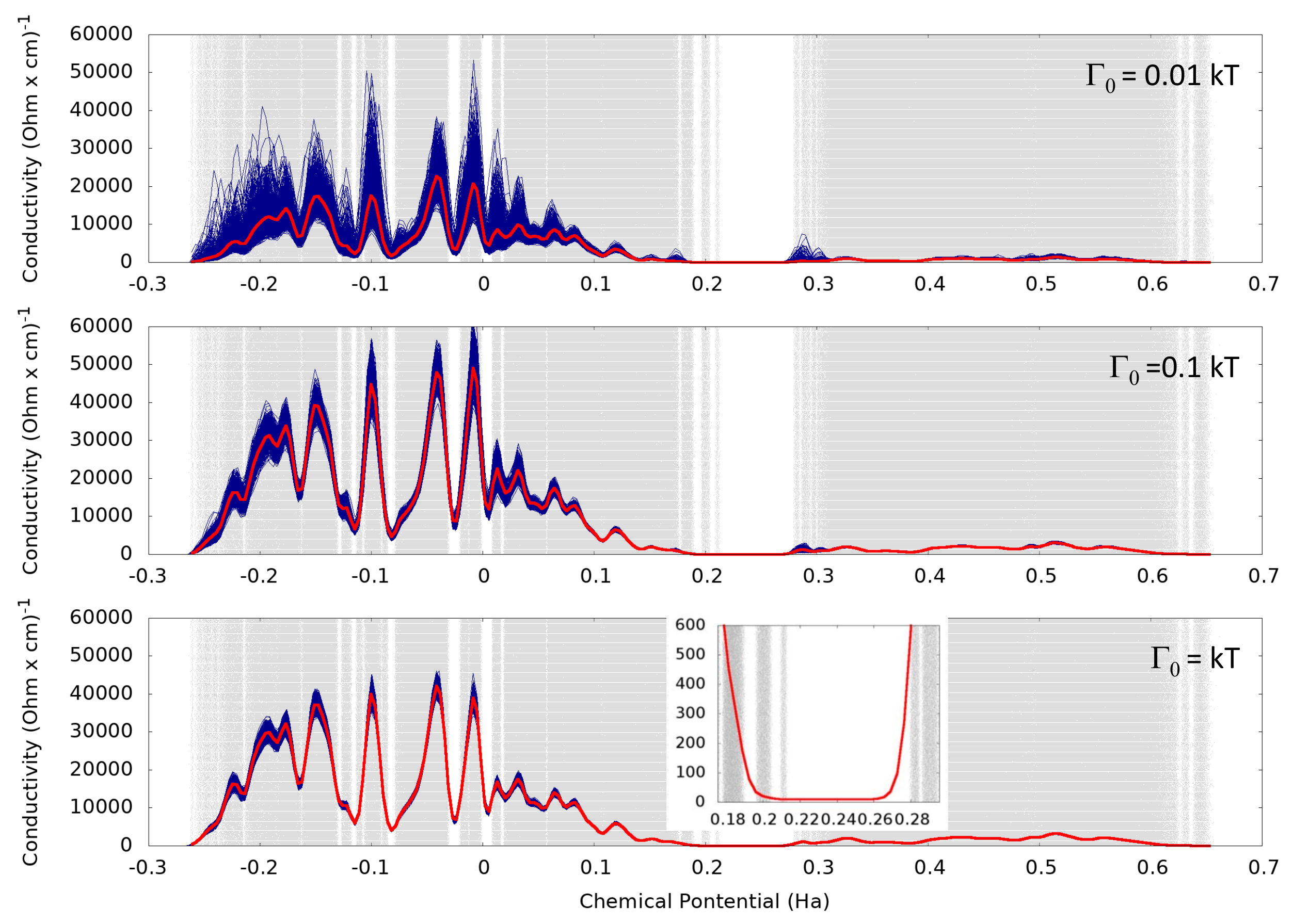}\\
  \caption{\scriptsize Same as Fig.~\ref{Fig:Sigma1000At300K} for $T=900$~K. In addition, the bottom panel contains a zoom-in.}
  \label{Fig:Sigma1000At900K}
\end{figure}

\subsection{Direct conductivity and mobilities of charge carriers}

\vspace{0.2cm}

In this section we present and analyze the numerical results on the direct conductivity \eqref{Eq:DCond}. Figs.~\ref{Fig:Sigma216At300K} and \ref{Fig:Sigma216At900K} report the data for a 216 Si atoms crystal at temperatures $T=300$~K and $T=900$~K, respectively. Similarly, Figs.~\ref{Fig:Sigma1000At300K} and \ref{Fig:Sigma1000At900K} report the data for a 1000 Si atoms crystal at temperatures $T=300$~K and $T=900$~K, respectively. The chemical potential has been varied throughout the entire energy spectral range and the dissipation parameter $\Gamma_0$ was sampled at three different values in these simulations. Overall, the results show a good correlation between the conductivity plots and the spectrum of the Kohn-Sham Hamiltonians. There is a significant difference between the outputs for 216 and 1000 atom crystals, indicating that the results are not converged yet w.r.t. the system's size. There is also a significant dependence on the dissipation parameter $\Gamma_0$. It is interesting to notice that for its largest values, the fluctuations of the direct conductivity are drastically suppressed. As explained in \cite{ProdanSpringer2017}, the convergence to thermodynamic limit is faster for larger $\Gamma_0$, and this explains the suppression of the fluctuations observed in these figures.

\vspace{0.2cm}

The conductivity results are in very good agreement with the mobility gap prediction, which for $T=900$~K can be found in Fig.~\ref{Fig:LevelStat}. Indeed, the conductivity is obviously not influenced at all by the first band of spectrum, which was determined in Fig.~\ref{Fig:LevelStat} to be Anderson localized. This is quite obvious in the inset of Fig.~\ref{Fig:Sigma1000At900K}, which shows a zoom into the region around the spectral gap. Let us point out again that, on the other hand, the intrinsic and extrinsic charge carriers are highly influenced by the presence of this Anderson localized band.

\vspace{0.2cm}

\begin{figure}
\center
  \includegraphics[width=0.9\textwidth]{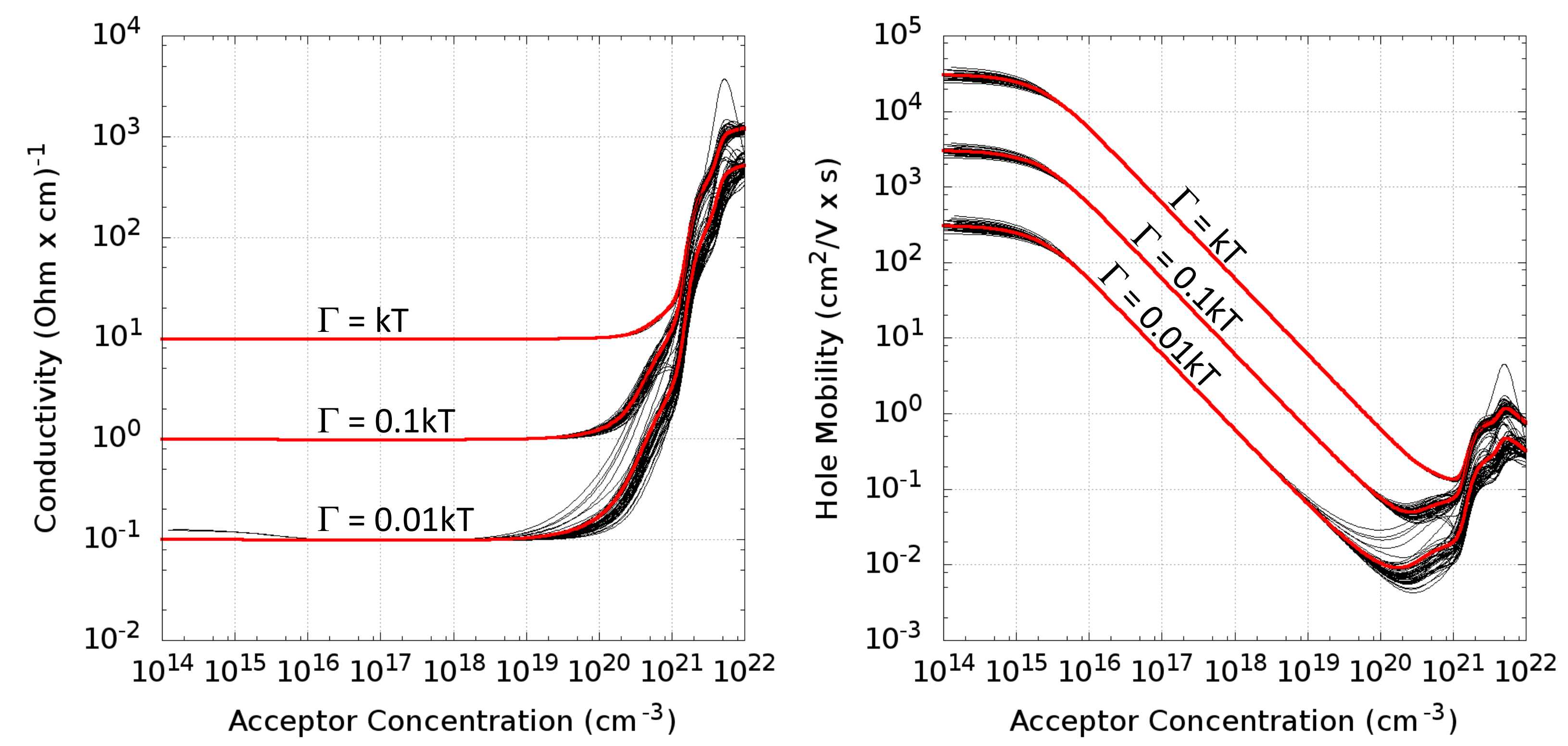}\\
  \caption{\scriptsize Dependence with the acceptor concentration of (a) direct conductivity and (b) hole mobility, as computed for the 1000-atom Si crystal at $T=900$~K. The fine lines represent unprocessed data coming from individual disordered configurations, hence they are a measure of the fluctuations. The thicker red lines represent the averages over 50 configurations. The simulations have been carried for three values of the dissipation parameter, which are specified in each panel.}
  \label{Fig:Zoom1000At900K}
\end{figure}

In Fig.~\ref{Fig:Zoom1000At900K}(a), we focus on the behavior of direct conductivity inside and around the insulating gap, especially on the hole side. We chose to investigate only at the crystal with $1000$ Si atoms and $T=900$~K because it is the most converged system. At $T=300$~K, the conductivity curves display a pronounced dependence on the spectral details which are not yet converged, hence the analysis will no be reliable. As it is customary, the transport coefficient has been plotted as a function of the acceptor concentration rather than chemical potential. The behavior of $\sigma$ seen in Fig.~\ref{Fig:Zoom1000At900K}(a) is as expected. Deep inside the insulating gap, the direct conductivity saturate at a value proportional to $\Gamma_0$ and, as the chemical potential moves towards the valence band, an activated behavior takes over. In Fig.~\ref{Fig:Zoom1000At900K}(b), we report the hole mobility as a function of the acceptor concentration. The functional shape is in good agreement with the measured one (see \cite[Fig.~21.8]{SpringerHandbook}). Let us point out that, when the chemical potential is inside a spectral gap, the mobility is proportional with the dissipation $\Gamma_0$, hence with inverse of the relaxation time. In contrast, when the chemical potential is inside a spectral band, as is the case of a metal, the mobility is proportional with relaxation time, hence inverse proportionally with $\Gamma_0$. As such, one should not be surprised by the behavior with $\Gamma_0$ seen in Fig.~\ref{Fig:Zoom1000At900K}(b).

\section{Conclusions}

We have derived disordered tight-binding models based on AIMD outputs and formulated a Kubo-formalism that preserves the self-averaging property of the transport coefficients. The Kubo-formalism was coded as a post-processing subroutine to a standard AIMD code and preliminary results on the transport coefficients of crystals Si were obtained at various temperatures. According to our study, the thermal disorder can have measurable effects even at room temperature.

\section{Acknowledgements} 

Emil Prodan acknowledges financial support from USA National Science Foundation through grant  DMR-1823800. Part of this project has received funding from the European Research Council (ERC) under the European Union's Horizon 2020 research and innovation programme (grant agreement No 716142). The authors would like to thank the Paderborn Center for Parallel Computing (PC$^2$) for the generous allocation of computing time on OCuLUS and the FPGA-based supercomputer NOCTUA.

\bibliographystyle{plain}

\end{document}